\documentclass[aps,prd,preprintnumbers,superscriptaddress,nofootinbib]{revtex4}% ,showpacs,twocolumn
\usepackage[dvips]{graphicx}
\usepackage{bm,latexsym,amsmath,amssymb,amsfonts,mathrsfs}
\usepackage{color}
\usepackage{subfigure}

\usepackage{remreset}
\input{colordvi.tex}

\newcommand{\be}{\begin{equation}}
\newcommand{\ee}{\end{equation}}
  \makeatletter

    	\@removefromreset{subtable}{chapter}
	
	\def\p@subtable{\arabic{table}}

  \makeatother
\begin{document}

\title{CMB distortion anisotropies due to the decay of primordial magnetic fields}

\author{Koichi~Miyamoto}
\email[Email: ]{miyamone"at"icrr.u-tokyo.ac.jp}
\affiliation{Institute for Cosmic Ray Research, The University of Tokyo,
Kashiwa, Chiba, 277-8582, Japan}

\author{Toyokazu~Sekiguchi}
\email[Email: ]{sekiguti"at"a.phys.nagoya-u.ac.jp}
\affiliation{Department of Physics, Nagoya University, Chikusa, Nagoya, 464-8602, Japan}
\affiliation{University of Helsinki and Helsinki Institute of Physics,
P.O. Box 64, FI-00014, Helsinki, Finland}

\author{Hiroyuki~Tashiro}
\email[Email: ]{hiroyuki.tashiro"at"asu.edu}
\affiliation{Physics Department, Arizona State University, Tempe, AZ 85287, USA}

\author{Shuichiro~Yokoyama}
\email[Email: ]{shu"at"icrr.u-tokyo.ac.jp}
\affiliation{Institute for Cosmic Ray Research, The University of Tokyo,
Kashiwa, Chiba, 277-8582, Japan}

\begin{abstract}
We investigate the power spectrum of the distortion of Cosmic Microwave Background (CMB)
due to the decay of the primordial magnetic fields.
It is known that there are two-types of the CMB distortions, so-called $\mu$- and $y$-types and
we find that the signal of the y-type distortion becomes larger than that of the $\mu$-type one.
We also discuss cross power spectra between the CMB distortions and the CMB temperature anisotropy, which are naturally generated due to the existence of the
primordial magnetic fields. We find that such cross power spectra have small amplitudes
compared with the
auto-power spectra of the CMB distortions because of the Silk damping effect
of the temperature anisotropy.
We also investigate the possibility of detecting such signal in the future CMB experiments,
including not only absolutely calibrated experiments such as PIXIE but also relatively calibrated experiments such as LiteBIRD and CMBpol.

\end{abstract}

\pacs{98.80.Cq}
\preprint{ICRR-Report-650-2012-39}
\maketitle
%%%%%%%%%%%%%%%%%%%%%%%%%%%%%%%%%%%%%%%%%%%%%%%%%%%%%%%%%%%%%%%%%%%%%%

\section{Introduction}

Recently, measurements of the Cosmic Microwave Background (CMB) spectral deviations 
from the black-body spectrum
have become a focus of attention as important probes of the physics in
the early Universe, because a powerful CMB observation missions called as
PIXIE and PRISM have been proposed~\cite{Kogut:2011xw, Andre:2013afa}.
Although the CMB spectrum is predicted as a nearly black-body
spectrum in the standard Big Bang scenario, spectral distortions from the black-body spectrum can be
created by energy injections into the CMB in the early universe.
Therefore, the measurement of CMB distortions is expected as a probe
of the thermal evolution of the Universe~(for recent reviews, see Refs.~\cite{Chluba:2011hw,Sunyaev:2013aoa}).
The diffusion of the acoustic waves before the recombination epoch, known as
Silk damping \cite{Silk:1967kq}, is one of the major energy injection
sources
\cite{1991MNRAS.248...52B,1991ApJ...371...14D,Hu:1994bz,Chluba:2012gq,Chluba:2012we,Dent:2012ne,Chluba:2011hw,Khatri:2012tv}.
Other
energy injection sources include massive unstable relic particles which
decay before the recombination epoch \cite{Hu:1993gc},
Hawking radiation from primordial black
holes \cite{Tashiro:2008sf}, diffusion damping of acoustic wave due to the cosmic strings~\cite{Tashiro:2012pp,Tashiro:2012nb},
and dissipation of
primordial magnetic fields before and after the recombination epoch~\cite{Jedamzik:1999bm, Sethi:2004pe, Kunze:2013uja}.

The CMB distortions are typically classified into two types,
so-called $\mu$- and $y$-distortions, depending on the epoch when
energy injections occur.  The $\mu$-distortions
are produced due to energy injections to CMB photons in the redshift
range $2 \times 10^6 \gtrsim z \gtrsim 5 \times 10^4$.  On the other
hand, the $y$-distortions are created by energy injections in the
redshift range $5 \times 10^4 \gtrsim z \gtrsim 1090$, and are also produced through the cosmic reionization
process \cite{Hu:1993tc} and the thermal Sunyaev-Zel'dovich(SZ) effect \cite{Zeldovich:1969ff}
from the clusters of galaxies \cite{Refregier:2000xz}.
Current constraints on these distortions have been respectively obtained as $|\mu| < 9 \times 10^{-5}$
and $y < 1.5 \times 10^{-5}$  from COBE FIRAS~\cite{Fixsen:1996nj}.
The future mission PIXIE has the potential to give 
tighter constraints on both
types of distortions, $|\mu| \sim5 \times 10^{-8}$ and
$y \sim 10^{-8}$ at the 5 $\sigma$ level \cite{Kogut:2011xw}, 
which will be improved further by an order of magnitude by another
future survey PRISM. 

In this paper, we investigate CMB distortions created by
energy injections due to the damping of the primordial magnetic fields.
Primordial magnetic fields could be the
seed fields of observed micro-Gauss magnetic fields in the
galaxies and galaxy clusters. 
There are a large number of works to study the origin of primordial magnetic fields in the early Universe;
during inflation (see, e.g., \cite{Ratra:1991bn,Martin:2007ue,Demozzi:2009fu} and references therein) or at the phase transition
(see, e.g., \cite{Hogan:1983zz,Vachaspati:1991nm,Enqvist:1994rm,Sigl:1996dm,Kahniashvili:2012uj} and references therein).
 Current upper limits on the large-scale
magnetic fields are obtained through CMB anisotropies (see, e.g., \cite{Shaw:2010ea,Shiraishi:2012rm,Yamazaki:2012pg})
and large scale structures (see, e.g., \cite{Shaw:2010ea, Pandey:2012ss, Kahniashvili:2012dy}). These upper limits allow
the existence of the nano-Gauss primordial magnetic fields on Mpc
scales. Recently, there are also several reports on the lower limits of magnetic fields in
the inter-galactic medium whose strength is larger than $O(10^{-15} -
10^{-20})$~Gauss by using the observations of TeV
blazars~\cite{Tavecchio:2010mk,Neronov:1900zz,Dolag:2010ni,Takahashi:2011ac}
although this claim is still under discussion \cite{Broderick:2011av, Miniati:2012ge}.

The effect of primordial magnetic fields on the CMB distortions
has been studied in Refs.~\cite{Jedamzik:1999bm, Sethi:2004pe, Kunze:2013uja}.  If primordial
magnetic fields exist, they induce the velocity of the photon-baryon fluid through
the Lorentz force before the recombination epoch. The induced kinetic energy dissipates 
through the viscosity of the photon-baryon fluid corresponding to
the Silk damping~\cite{Jedamzik:1996wp,Subramanian:1997gi}.
Even after the recombination, the magnetic fields
induce the velocity of baryon fluid via the Lorentz force with residual ionized baryons.
This velocity fields also dissipate by  
ambipolar diffusion and decaying magnetohydrodynamical turbulence,
and, consequently, CMB distortions are produced~\cite{Sethi:2004pe, Kunze:2013uja}.
For example, calculating spatially averaged distortions due to the
magnetic field damping before the recombination epoch, 
the authors of Ref.~\cite{Jedamzik:1996wp}
have obtained the upper limits on the strength of the
magnetic fields by comparing the results from COBE-FIRAS, which
are $ 3\times 10^{-8} $ Gauss on comoving coherent scale $\sim 400~{\rm
pc}$ from the constraint for $\mu$-distortions ($0.3~{\rm pc}$ for $y$-distortions).
Recently, in Ref.~\cite{Kunze:2013uja},
the authors have claimed that the PIXIE would be expected to give a constraint as $8 \times 10^{-10}$ Gauss
from the limit on $|\mu|$.

In this paper, we focus on the anisotropies of CMB
distortions induced by primordial magnetic fields.
In the future experiments,
it is expected 
to measure such anisotropies of the distortion before the recombination epoch.  We investigate the
angular power spectrum of the $\mu$- and $y$-distortions due to the
damping of primordial magnetic fields with a given initial power
spectrum.
The shape of the angular power spectrum, in particular, the existence of
the peak of the spectrum, is expected to depend on the kind of energy injections.
We show that 
the amplitude of the spectrum depends on the structure of primordial
magnetic fields and the peak scale informs us about the dissipation scale
of magnetic fields.

We
also evaluate the cross-correlation between the CMB distortion and
the CMB temperature anisotropies.  There are several works about such
cross-correlation in the context of searching primordial
non-Gaussianity~\cite{Pajer:2012vz,Ganc:2012ae}.
If the magnetic fields exist,
for example,
these fields generate the anisotropic
stress during the radiation-dominated era which becomes a source of the
additional primordial curvature perturbations.
CMB temperature fluctuations induced by such primordial curvature perturbations
sourced from the anisotropic stress of primordial magnetic fields
would correlate with the CMB distortions due to the damping of primordial magnetic fields,
because both of them are given in terms of the convolution of the magnetic fields.
Including the analysis of such cross-correlation, we discuss the
possibility of detecting the CMB distortions due to the existence of primordial magnetic fields.
 
This paper is organized as follows.
In section 2, 
we briefly review CMB distortions induced from the damping of the magnetic fields and
present the formalism for calculation of angular power spectra of anisotropies of $\mu$ and $y$ parameters.
We also discuss the cross-correlation between the CMB distortions and the CMB temperature anisotropy
induced from the primordial magnetic fields.
In section 3, we numerically calculate angular power spectra of the CMB distortions, taking the amplitude of the
primordial magnetic fields to be a largest possible one derived from the current CMB observations.
In section 4, we discuss the possibility of detecting anisotropic $\mu$- and $y$-distortions in future or on-going CMB experiments.
In section 5, we conclude this paper.

In this paper, we use the natural unit: $\hbar=c=k_B=1$.
Cosmological parameters are set according to WMAP result\cite{Hinshaw:2012fq}: the abundance of baryon $\Omega_b=0.045$,
that of cold dark matter $\Omega_c=0.222$, that of dark energy $\Omega_\Lambda=0.733$ and Hubble constant $H_0=70.4~{\rm km/s/Mpc}$.

\section{Formulation for CMB distortions due to primordial magnetic fields}

\subsection{Primordial magnetic fields}

We assume that spatially-varying random magnetic fields
$\mathbf{B}(z,\mathbf{x})$ are created in the early universe.
We define $\mathbf{b}(z,\mathbf{x})$ as 
\be
\mathbf{B}(z,\mathbf{x})=\frac{\mathbf{b}(z,\mathbf{x})}{a^2}, 
\ee 
where
$a$ is the scale factor,
and $\mathbf{b}(z,\mathbf{x})$ describes
the evolution of magnetic fields other than decay due to cosmic expansion.
In addition to the cosmic expansion, small-scale magnetic fields lose their
amplitude 
through the dissipation process due to the viscous photon-baryon
fluid before the recombination epoch
\cite{Jedamzik:1996wp}.
Accordingly, the time-evolution of $\tilde{\mathbf{b}}(z,\mathbf{k})$,
which is the Fourier transformed component of $\mathbf{b}(z,\mathbf{x})$ with comoving wavenumber
$\mathbf{k}$, is given by
\be
\mathbf{\tilde{b}}(z,\mathbf{k})=\mathbf{\tilde{b}}(\mathbf{k})\exp(-\tau(z,\mathbf{k})),
\ee
where 
\be \tau(z,\mathbf{k})=-\int^{t(z)}_{t(z_0)}dt^{\prime}
~\Gamma (t^\prime,\mathbf{k}),
\ee
with the dissipation rate $\Gamma (t,\mathbf{k})$.
Here, we take $z=z_0$ to be an arbitrary initial redshift
when magnetic fields on interesting scales have hardly decayed yet
and $\mathbf{\tilde{b}}(\mathbf{k}) = \tilde{\mathbf{b}}(z_0,\mathbf{k}) $.
Note that $\tau(z,k)>1$ means that magnetic fields with wavenumber $k$ have almost decayed at redshift $z$.

We assume that the initial random magnetic fields are
isotropically homogeneous and obey the Gaussian statistics.
Therefore, the auto-correlation function of $\mathbf{\tilde{b}}(\mathbf{k})$ is expressed as
\be
\left< \tilde{b}_i(\mathbf{k})\tilde{b}_j(\mathbf{p})\right>=P_{ij}(\hat k)P_B(k)(2\pi)^3\delta(\mathbf{k}+\mathbf{p}),
\label{magauto}
\ee
where $k=|\mathbf{k}|$, $p=|\mathbf{p}|$ and 
\be
P_{ij}(\hat k)=\delta_{ij}-\hat{k}_i\hat{k}_j,
\ee
is a projection tensor which reflects the zero divergence of magnetic fields.
We assume that the power spectrum, $P_B$, is given as a blue-tilted power-law function with a cut-off scale, defined as
\be
P_B(k)=
\begin{cases}
n\pi^2\frac{B_0^2}{k^3}\left(\frac{k}{k_c}\right)^n& \ ;k<k_c \\
0& \ ;k>k_c 
\end{cases},
\label{P_B}
\ee
where $n > 0 $ is the spectral index \footnote{ Although, here, we do not mention concrete models of generating the primordial magnetic fields,
such blue-tilted power spectrum is motivated by some models, e.g., the phase transition scenarios
in the early universe \cite{Hogan:1983zz,Vachaspati:1991nm,Enqvist:1994rm,Sigl:1996dm,Kahniashvili:2012uj}.}
and 
$k_c$ is the cut-off wavenumber depending on the generation mechanism of the magnetic fields.

The dissipation rate $\Gamma (t,\mathbf{k})$ is expressed as the
imaginary part in the solutions of dispersion relations for the magnetohydrodynamic (MHD)
modes, called fast- and slow-magnetosonic, and Alfven modes. 
In Ref.~\cite{Jedamzik:1996wp}, the authors have shown that, among these modes, the Alfven and slow-magnetosonic modes
can survive below the Silk damping scale. Therefore the
energy of the magnetic fields can be stored in these modes and dissipate 
with damping rates of 
the Alfven and slow-magnetosonic modes, which depend on scales.
On the scale larger than the mean free path for photon
$l_\gamma$ , i.e., $k/a \lesssim l_\gamma^{-1}$,
the damping of MHD modes is caused by the photon shear viscosity.
On the other hand, on the scale smaller than $l_\gamma$, i.e., $k/a \gtrsim l_\gamma^{-1}$, MHD modes are damped by the occasional collisions of
the fluid particles with the background ones, which is parametrized by the drag coefficient 
$\alpha\simeq (l_\gamma R)^{-1}$ with $R={3\rho_b \over 4\rho_r}$ being the ratio between the energy densities of the baryon $\rho_b$ and the radiation $\rho_r$.
Furthermore, the damping rate is different in the oscillatory limit and the overdamped limit
and then 
the dissipation rate, depending on the scales, is obtained as \cite{Jedamzik:1996wp}
\be \Gamma (t,\mathbf{k})\sim
\begin{cases}
0 & ;  {\rm for} 
\ \frac{k}{a}\lesssim H 
~~~{\rm (no~damping~for~superHubble~modes)}\\
\frac{l_\gamma }{10 (1+R)}\left(\frac{k}{a}\right)^2 &  ;  {\rm for} 
\ H \lesssim \frac{k}{a} \lesssim \frac{30 v_A\cos\theta(1+R)}{l_\gamma } 
~~~{\rm (oscillatory~limit~for~photon~shear~viscosity)}\\
\frac{v_A^2\cos^2\theta}{5 l_\gamma } & ;{\rm for} 
\ \frac{30 v_A\cos\theta(1+R)}{l_\gamma}  \lesssim  \frac{k}{a} \lesssim l_\gamma^{-1} 
~~~ {\rm (overdamped~limit~for~photon~shear~viscosity)}\\
\frac{c_A^2\cos^2\theta}{\alpha}\left(\frac{k}{a}\right)^2 & ;{\rm for} 
\  l_\gamma^{-1} \lesssim \frac{k}{a} \lesssim \frac{\alpha}{2c_A\cos\theta} 
~~~ {\rm (overdamped~limit~for~occasional~collisions)}\\
\frac{\alpha}{2} & ;{\rm for} 
\ \frac{k}{a} \gtrsim \frac{\alpha}{2c_A\cos\theta}
~~{~\rm (oscillatory~limit~for~occasional~collisions)}
\end{cases},
\label{Imomega}
\ee
with
\be
v_A^2 = {\hat{\mathbf{B}}_{\rm eff}^2 \over (1+R+\hat{\mathbf{B}}_{\rm eff}^2)},
~~
c_A^2 = {\hat{\mathbf{B}}_{\rm eff}^2 \over R}.
\ee
Here
$H$ is the Hubble parameter, $\theta$ is the angle between $\hat{\mathbf{B}}_{\rm eff}$ and
$\mathbf{k}$,
$v_A$ and $c_A$ respectively denote
the relativistic and non-relativistic Alfven velocities \cite{Jedamzik:1996wp,Seshadri:2000ky,Mack:2001gc}
and the normalized mean square of the effective background field
$\hat{\mathbf{B}}_{\rm eff}$ is given by
\be
\hat{\mathbf{B}}_{\rm eff}^2 \equiv
{\mathbf{B}_{\rm eff}^2 \over 16 \pi \rho_r / 3}
={ \left< \mathbf{B}^2(z,\mathbf{x}) \right>
\over 16 \pi \rho_r / 3}
={3 \over 16 \pi \rho_r} \int \frac{dk}{\pi^2}k^2P_B(k)\frac{1}{a^4}e^{-2\tau(z,k)}.
\ee
We hereafter simply set as $\cos \theta=1$, and regard $\tau(z,\mathbf{k})$ as the function of $k$.

\subsection{Auto- and cross-correlation functions of $\mu$ and $y$ parameters}

The dissipation of primordial magnetic fields discussed above can
be a mechanism of energy injection which 
creates the CMB spectral distortions.
Since the amplitude of magnetic fields spatially varies, the dissipation
energy of the magnetic fields also spatially fluctuates and 
it can produce 
the anisotropic spectral distortions of the CMB.

The spectral distortions of the CMB are characterized by $\mu$ and $y$
parameters. 
These parameters are given by~\cite{Sunyaev:1980vz,Hu:1992dc} \footnote{Recently, Refs.~\cite{Chluba:2012gq,Kunze:2013uja} found that
an extra factor $1/3$ in the Eqs.~(\ref{mu_general}) and (\ref{y_general}) is needed because only $1/3$ of the energy injection contributes to the distortions. However this modification do not change our final results significantly.}
\be
\mu(\mathbf{x})=1.4 \int^{z_{\mu,i}}_{z_{\mu,f}} dz \frac{dQ(z,\mathbf{x})/dz}{\rho_\gamma(z)} ,\label{mu_general}
\ee
and
\be
y(\mathbf{x})=\frac{1}{4} \int^{z_{y,i}}_{z_{y,f}} dz 
\frac{dQ(z,\mathbf{x})/dz}{\rho_\gamma(z)}, \label{y_general}
\ee
respectively.
Here, $dQ(z,\mathbf{x})/dz$ is the energy injected at redshift $z$ and
comoving coordinate $\mathbf{x}$, $\rho_\gamma(z)$ is the photon energy
density, and we take
$z_{\mu,i}= 2\times 10^6$, $z_{\mu,f}=z_{y,i}= 5\times 10^4$ and $z_{y,f} =z_{\rm rec}= 1090$.
$z_{\rm rec}$ is also the redshift at the recombination.

The injected energy is given by~\cite{Jedamzik:1999bm}
\be
\frac{dQ}{dz}(z,\mathbf{x})=-\frac{1}{8\pi a^4}\frac{d}{dz}\left(\mathbf{b}(z,\mathbf{x})\right)^2. \label{dQdz}
\ee
Substituting Eq. (\ref{dQdz}) into Eqs.~(\ref{mu_general}) and (\ref{y_general}),
$\mu$ and $y$ parameters induced by dissipating magnetic fields are respectively given by
\be
\mu(\mathbf{x})=\frac{1.4}{8\pi}\left(\frac{(\mathbf{b}(z_{\mu,i},\mathbf{x}))^2}{\rho_{\gamma,0}}-
\frac{(\mathbf{b}(z_{\mu,f},\mathbf{x}))^2}{\rho_{\gamma,0}}\right)
\label{mu_B},
\ee
\be
y(\mathbf{x})=\frac{1}{32\pi}\left(\frac{(\mathbf{b}(z_{y,i},\mathbf{x}))^2}{\rho_{\gamma,0}}-\frac{(\mathbf{b}(z_{y,f},\mathbf{x}))^2}{\rho_{\gamma,0}}\right)
\label{y_B},
\ee
where $\rho_{\gamma,0}=\rho_\gamma(0)$.
In terms of the Fourier modes of magnetic fields, $\tilde{\mathbf{b}}(z,\mathbf{k})$, we can rewrite these
parameters as
\be
\mu(\mathbf{x})=\frac{1.4}{8\pi \rho_{\gamma,0}}\int 
\frac{d^3k}{(2\pi)^3}\int \frac{d^3k^\prime}{(2\pi)^3} ~
\tilde{\mathbf{b}}(\mathbf{k})\cdot\tilde{\mathbf{b}}^*(\mathbf{k}^\prime)C_\mu(k,k^\prime)e^{i(\mathbf{k}-\mathbf{k}^\prime)\cdot \mathbf{x}},
\label{mux}
\ee
\be
y(\mathbf{x})=\frac{1}{32\pi \rho_{\gamma,0}}\int
\frac{d^3k}{(2\pi)^3}\int \frac{d^3k^\prime}{(2\pi)^3} ~
 \tilde{\mathbf{b}}(\mathbf{k})\cdot\tilde{\mathbf{b}}^*(\mathbf{k}^\prime)C_y(k,k^\prime)e^{i(\mathbf{k}-\mathbf{k}^\prime)\cdot \mathbf{x}},
 \label{yx}
\ee
where
\be
C_\mu(k,k^\prime)=\exp\left(-\tau(z_{\mu,i},k)\right)\exp\left(-\tau(z_{\mu,i},k^{\prime})\right)-\exp\left(-\tau(z_{\mu,f},k)\right)\exp\left(-\tau(z_{\mu,f},k^{\prime})\right),
\label{Cmu}
\ee
\be
C_y(k,k^\prime)=\exp\left(-\tau(z_{y,i},k)\right)\exp\left(-\tau(z_{y,i},k^{\prime})\right)-\exp\left(-\tau(z_{y,f},k)\right)\exp\left(-\tau(z_{y,f},k^{\prime})\right).
\label{Cy}
\ee

Let us discuss the angular power spectrum of the distortions.
First, considering the expansion of the distortion parameters, $\mu$ and $y$, by
the spherical harmonics, $Y_{lm}(\hat{n})$, we can obtain each mode-coefficient as
\be
a^{\mu}_{lm} =\int d^2\hat{n}~ \mu(r_{\rm rec}\hat{n})Y^*_{lm}(\hat{n}),
\label{eq:alm_mu}
\ee
and
\be
a^{y}_{lm} =\int d^2\hat{n}~ y(r_{\rm rec}\hat{n})Y^*_{lm}(\hat{n}) ,
\label{eq:alm_yy}
\ee
where we take the sudden last-scattering approximation in which
the observed CMB photons are last-scattered simultaneously at $z=z_{\rm rec}$.
In Eqs.~(\ref{eq:alm_mu}) and (\ref{eq:alm_yy}),
$\hat{n}$ is the direction of the line of sight and $r_{\rm
rec}=\int^{z_{\rm rec}}_0 dz/H(z)\simeq 1.4\times 10^4 ~{\rm Mpc}$
is the comoving distance from the earth to the last-scattering surface.
Angular power spectra of two kinds of distortions are given by
\be
\left< a^X_{lm}  (a^Y_{l^\prime m^\prime})^*\right> = C^{XY}_l \delta_{ll^\prime}\delta_{mm^\prime},
\ee
where $X$ and $Y$ are either $\mu$ or $y$.
According to Eqs.~(\ref{magauto}), (\ref{mux}), (\ref{yx}), (\ref{eq:alm_mu}) and (\ref{eq:alm_yy}),
we have
\be
C^{\mu\mu}_l=\frac{1.4^2}{2(2\pi)^5\rho_{\gamma,0}^2}\int dp\int
dq\int^1_{-1}d\mu ~
p^2q^2 P_B(\chi)P_B(q)\left(C_\mu(\chi,q)\right)^2f(p,q,\mu)\left(j_l(pr_{\rm rec})\right)^2,
\label{CmumuB}
\ee
\be
C^{yy}_l=\frac{1}{32(2\pi)^5\rho_{\gamma,0}^2}\int dp\int
dq\int^1_{-1}d\mu~
p^2q^2 P_B(\chi)P_B(q)\left(C_y(\chi,q)\right)^2f(p,q,\mu)\left(j_l(pr_{\rm rec})\right)^2,
\label{CyyB}
\ee
and
\be
C^{\mu y}_l=\frac{1.4}{8(2\pi)^5\rho_{\gamma,0}^2}\int dp\int
dq\int^1_{-1}d\mu~
p^2q^2 P_B(\chi)P_B(q)C_\mu(\chi,q)C_y(\chi,q)f(p,q,\mu)\left(j_l(pr_{\rm rec})\right)^2,
\label{CmuyB}
\ee
where 
\be
\chi=\sqrt{p^2+q^2+2pq\mu},
 \quad
f(p,q,\mu)=\frac{p^2(1+\mu^2)+4pq\mu+2q^2}{p^2+2pq\mu+q^2},
\ee
and $j_l$ is the $l$-th spherical Bessel function.

Note that
the finite thickness of the last scattering surface cannot be neglected for smaller scale anisotropies $l\gtrsim 1000$
and the sudden last-scattering approximation is not valid for such scales.
According to Ref.~\cite{Subramanian:1998fn}, however,
the effect of the finite thickness of the last scattering surface can be simply taken into account in the above expressions as
\be
C^{XY}_l\approx
\begin{cases}
C^{XY,0}_l \ {\rm ;for} \ l<r_{\rm rec}/\sigma_{\rm LS} \\
\frac{C^{XY,0}_l}{l\sigma_{\rm LS}/r_{\rm rec}} \ {\rm ;for} \ l>r_{\rm rec}/\sigma_{\rm LS}
\end{cases},
\label{finiteLS}
\ee
where $C^{XY,0}_l$ is the angular power spectrum
given by Eq. (\ref{CmumuB}), (\ref{CyyB}) or (\ref{CmuyB})
and $\sigma_{\rm LS}\simeq 17~{\rm Mpc}$~\cite{Subramanian:2002nh} is the thickness of the last scattering surface.

\subsection{Cross-correlation functions between CMB distortions and CMB temperature anisotropies}

The cross correlation between the CMB temperature
anisotropy and the CMB spectral distortion is exactly zero as long as
the primordial curvature perturbations are pure Gaussian, and hence it
would be a new probe of the non-Gaussian feature of the primordial curvature perturbations \cite{Pajer:2012vz,Ganc:2012ae}.
Primordial magnetic fields generate not only CMB $\mu$ and $y$ distortions as shown in the previous
subsection, but also
the large-scale temperature anisotropy of
the CMB \cite{Shaw:2010ea,Shiraishi:2012rm,Yamazaki:2012pg},
which are in general given as a quadratic function of the random Gaussian magnetic fields, $\mathbf{B}$, as shown below.
Since the $\mu$ and $y$ parameters are also proportional to $\mathbf{B}^2$ as shown in Eqs. (\ref{mu_B}) and (\ref{y_B}),
the primordial magnetic fields can make
non-zero cross-correlation between the CMB temperature and spectral
distortion anisotropies.
Therefore, we investigate such a
cross-correlation as a signature of primordial magnetic fields in this section.

One of the effects of the primordial magnetic fields on the CMB temperature anisotropy
is so-called a scalar passive mode,
\footnote{There is also another type of CMB fluctuations called the scalar magnetic mode\cite{Shaw:2009nf}.
We will discuss this mode later and show that both of the cross-correlation angular power spectra
due to the scalar passive mode and the scalar magnetic mode are far below the detectable level in future experiments in following sections.
The cross-correlation angular power spectrum
between the vector or tensor mode of the temperature anisotropy and the
$\mu$ or $y$ anisotropy vanishes, since $\mu$ and $y$ are scalar-like quantities.
}
which is extra curvature perturbations induced from the magnetic anisotropic stress
on super-horizon scales
generated during radiation dominated era before the neutrino decoupling time.
The scalar passive mode of the
curvature perturbations on the comoving slicing, $\zeta_{sp}$, is given by~\cite{Shaw:2009nf}
\be
\zeta_{sp}(\mathbf{k})=-\frac{1}{3}R_\gamma\Pi_B(\mathbf{k})\left(\ln\left(\frac{\eta_\nu}{\eta_B}\right)+\frac{5}{8R_\nu}-1\right),\label{zetasp}
\ee
where $\Pi_B$ is the scalar part of the anisotropic stress of
magnetic fields, $R_\gamma=\rho_\gamma/\rho_r$,
$\rho_r=\rho_\gamma+\rho_\nu$ is the energy density of relativistic
particles, $\rho_\nu$ is the neutrino energy density,
$R_\nu=\rho_\nu/\rho_r$, $\eta_\nu$ is the conformal time at neutrino
decoupling and $\eta_B$ is that at magnetic field generation.  
We
hereafter set $\eta_\nu/\eta_B=10^{17}$.
This value corresponds to magnetic fields generated at the energy scale of Grand Unified Theory and maximizes the scalar passive mode.
The scalar part of the anisotropic stress, $\Pi_B$, is given by
\be
\Pi_B(\mathbf{k})=\frac{9}{2}T_{ij}(\hat{k})\Delta^{ij}(\mathbf{k}),
\ee
where
\be \Delta^{ij}(\mathbf{k})=\frac{1}{4\pi\rho_{\gamma,0}}\int
\frac{d^3p}{(2\pi)^3}\int
\frac{d^3q}{(2\pi)^3}\tilde{b}^i(\mathbf{p})\tilde{b}^j(\mathbf{q})(2\pi)^3\delta(\mathbf{k}-\mathbf{p}-\mathbf{q}),
\ee
and
\be
T_{ij}(\hat{k})=\hat{k}_i\hat{k}_j-\frac{1}{3}\delta_{ij}.
\ee
The multipole coefficient of the scalar passive mode 
is given in terms of $\zeta_{sp}$ as
\be
a^{T,sp}_{lm}=4\pi
i^l\int\frac{d^3k}{(2\pi)^3}
\Delta^S_l(k)\zeta_{sp}(\mathbf{k})Y^*_{lm}(\hat k), \label{alm_Tsp}
\ee
where $\Delta^S_l(k)$ is the transfer function of the scalar mode which
we calculate using CAMB~\cite{Lewis:1999bs,CAMBsite}.

Then we can compute the cross-correlation angular power spectra between the CMB distortions and the CMB temperature anisotropies, which are defined as
\be
\left< a^X_{lm}  (a^T_{l^\prime m^\prime})^*\right>=C^{XT}_l \delta_{ll^\prime}\delta_{mm^\prime},
\ee
where $X$ is $\mu$ or $y$ again.
From Eqs~(\ref{magauto}), (\ref{mux}), (\ref{yx}), (\ref{eq:alm_mu}), (\ref{eq:alm_yy}), (\ref{zetasp}) and (\ref{alm_Tsp}),
explicit forms of $C^{XT}_l$ are given by
\be
C^{\mu T}_l=1.4A\int dp\int dq \int ^1_{-1}d\mu p^2q^2P_B(\chi)P_B(q)C_\mu(q,\chi)\Delta^S_l(p)g(p,q,\mu)j_l(pr_{LS}),
\label{CmuTB}
\ee
and
\be
C^{yT}_l=\frac{1}{4}A\int dp\int dq \int ^1_{-1}d\mu p^2q^2P_B(\chi)P_B(q)C_y(q,\chi)\Delta^S_l(p)g(p,q,\mu)j_l(pr_{LS}),
\label{CyTB}
\ee
where 
\be
g(p,q,\mu)=\frac{(1-3\mu^2)k^2-(1+\mu^2)p^2-(1+3\mu^2)kp\mu}{3(k^2+p^2+2kp\mu)},
\ee
and $A=\frac{1}{(2\pi)^5}\frac{3R_\gamma}{2\rho_{\gamma,0}^2}\left(\ln\left(\frac{\eta_\nu}{\eta_B}\right)+\frac{5}{8R_\nu}-1\right)$.

\section{Estimate of power spectra of the CMB distortion parameters}

\subsection{Current upper limit on primordial magnetic fields and scale of magnetic field decay}

\begin{figure}[tbp]
\begin{tabular}{cc}
\begin{minipage}{0.5\hsize}
\begin{center}
  \subfigure[$\tau(z_{\mu,f},k)$]{
\includegraphics[width=75mm]
{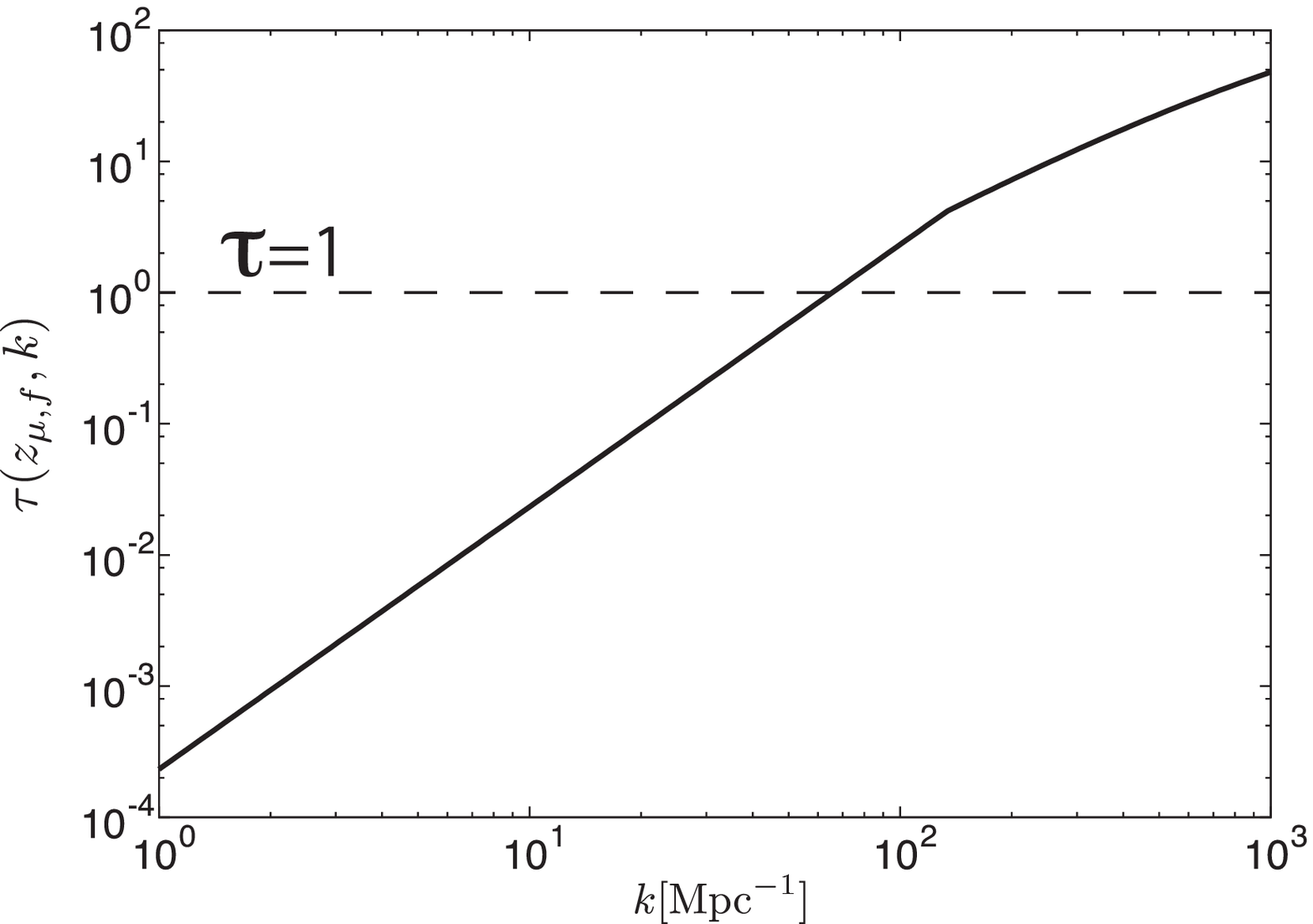}
\label{fig:tau_mu}
}
\end{center}
\end{minipage}

\begin{minipage}{0.5\hsize}
\begin{center}
  \subfigure[$\tau(z_{y,f},k)$]{
\includegraphics[width=75mm]
{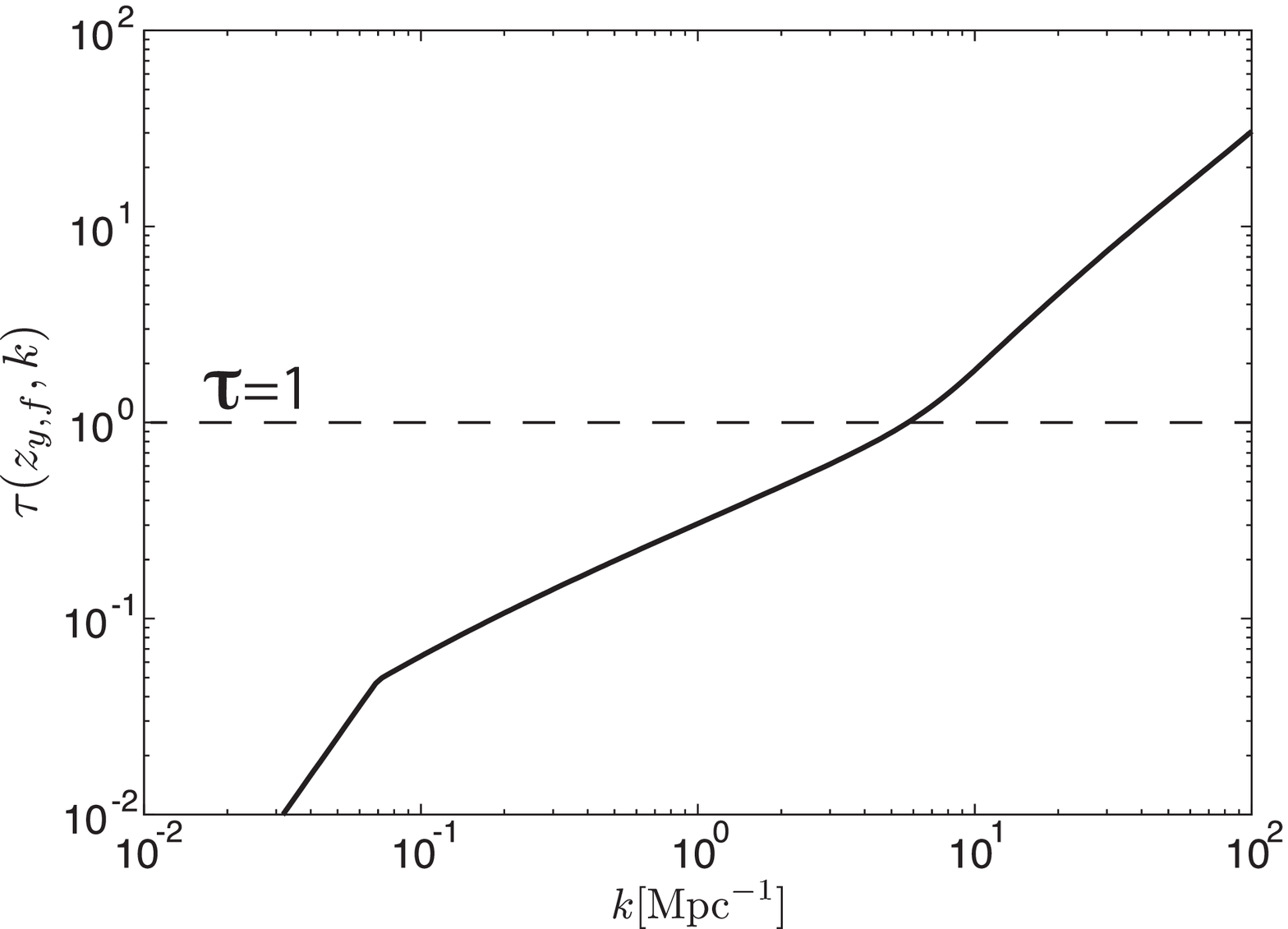}
\label{fig:tau_y}
}
\end{center}
\end{minipage}
\end{tabular}
\caption{The dependence of $\tau(z_{\mu,f},k)$ and $\tau(z_{y,f},k)$ on $k$.
In both figures, we plot $B_0=1.1\times 10^2\, {\rm nG}$, $n=4$ and $k_c=100\,{\rm Mpc}^{-1}$ for $\tau(z_{\mu,f},k)$ and $k_c=10\,{\rm Mpc}^{-1}$ for $\tau(z_{y,f},k)$. The dashed line in each figure represents $\tau=1$.
}
\label{fig:tau}
\end{figure}

In order to evaluate the angular power spectra of the CMB distortion anisotropies derived in the previous section,
we need to set  parameters $B_0$, $n$ and $k_c$, which specify the power spectrum of primordial magnetic fields.
In this subsection, we briefly review the current constraints on these parameters.
Then, in the next subsection,
we consider situations where observational signals of $\mu$ and $y$ anisotropies are maximized within such constraints.

One of strong cosmological constraints on primordial magnetic fields is
that obtained from the isotropic CMB distortion by COBE FIRAS~\cite{Fixsen:1996nj}: $|\mu| < 9 \times 10^{-5}$
and $y < 1.5 \times 10^{-5}$.
These limits bring the upper bound of the decaying energy density of magnetic
fields during the era when CMB distortions are created: $\rho_B=\mathbf{B}_{\rm
eff}^2/8\pi\lesssim10^{-4}\times\rho_\gamma$
~\cite{Jedamzik:1999bm}.
In terms of $B_0$,
this constraint leads 
\be
B_0<1.1\times 10^2~{\rm nG}, \label{COBEcons}
\ee
which is independent of $n$ and $k_c$, 
if $k_c$ is at the scale where magnetic fields decay while CMB distortions can be generated.

Another important constraint is that from observations of CMB temperature anisotropies.
Ref.~\cite{Yamazaki:2010nf} derived the upper limit of the amplitude of primordial magnetic fields
\be
|B_\lambda|<3.0{\rm nG}, \label{CMBcons}
\ee
where $B_\lambda$
corresponds to the strength of primordial magnetic fields on a comoving scale of $1\,{\rm Mpc}$,
which is related with $B_0$ as
\be
B_0=\left[\frac{2}{n\Gamma(n/2)}\right]^{1/2}(2\pi)^{n/2}\left(\frac{k_c}{k_\lambda}\right)^{n/2}B_\lambda,
\label{B0Blambda}
\ee
with $k_\lambda=2\pi\, {\rm Mpc}^{-1}$.
From the above expression, we find that the constraint for $B_0$ obtained from the CMB temperature anisotropy
depends on the spectral index $n$ and the cut-off wavenumber $k_c$.

\begin{figure}[tbp]
\begin{center}
\includegraphics[width=80mm]
{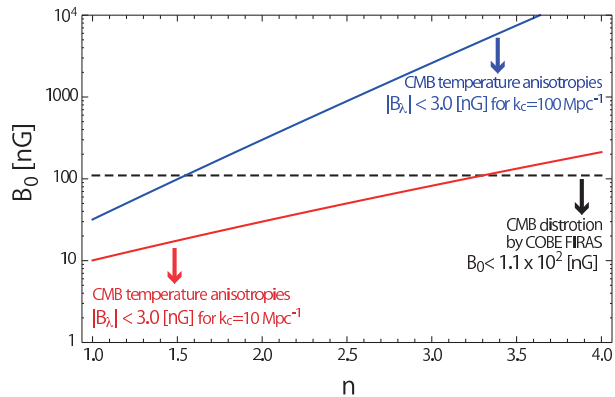}
\end{center}
\caption{The allowed region in the plane of $B_0$, the amplitude of the magnetic field power spectrum, and $n$, its tilt.
The black dashed line shows the upper bound given by Eq.~(\ref{COBEcons}), which is obtained from the observation of CMB distortion by COBE.
The blue (red) line shows the constraint Eq.~(\ref{CMBcons}) from current observations of CMB temperature anisotropies for $k_c=100 (10)~{\rm Mpc}^{-1}$.}
\label{fig:n-B0}
\end{figure}

Our aim of this paper is to evaluate the maximum signals of the CMB distortion
anisotropy due to primordial magnetic fields.
Basically, the power spectrum of the CMB distortions due to the decay of the magnetic fields has a peak at the cut-off scale ($\sim 1/k_c$)
and the peak amplitude depends on the total decaying energy density of
magnetic fields over all scales, not only on the peak scale. 
Hence in order to obtain the large amplitude on the observable scales, which are much larger than the peak scale,
we take the peak scale characterized by $k_c$ as large as possible (We discuss the
details in the next subsection). Since the typical scale of the decay of the magnetic fields becomes larger as the Universe expands,
we set the peak scale of the power spectrum to the scale
on which magnetic fields decay around the end of the production era of
CMB distortions.
This means that $k_c$ satisfies $\tau(z_{\mu,f}, k_c) \sim 1$
for the $\mu$-distortion and
$\tau(z_{y,f}, k_c) \sim 1$
for the $y$-distortion.
We plot $\tau(z_{\mu,f},k)$ and $\tau(z_{y,f},k)$ as functions of $k$ in FIG.~\ref{fig:tau}.
In both figures, we take
$B_0=1.1\times 10^2 {\rm nG}$ and $n=4$ which satisfy the COBE bound.
According to FIG.~\ref{fig:tau}, we set
$k_c=100\,{\rm Mpc}^{-1}$ for the $\mu$ distortion and $k_c=10\,{\rm Mpc}^{-1}$ for $y$ distortion.
Note that $\tau(z_{X,f},k)<1$ for $k < k_c$ means that magnetic fields hardly decay during the production
era of CMB distortions and $C^{XX}_l$ is strongly suppressed.
In FIG. \ref{fig:n-B0}, we show the region in the $n$-$B_0$ plane excluded by Eqs.~(\ref{COBEcons}) and (\ref{CMBcons}), and
we set the peak of the magnetic field power spectrum $k_c$ as $k_c=10\,{\rm Mpc}^{-1}$ or $k_c=100\,{\rm Mpc}^{-1}$ for the constraint given by Eq. (\ref{CMBcons}).
The upper limit on $B_0$ from Eq.~(\ref{COBEcons}) does not depends on $n$, which is because of out notation (\ref{P_B}). 
On the other hand, the constraint Eq.~(\ref{CMBcons}) becomes less severe as $n$ increases, since for large $n$ the magnetic field power spectrum is highly peaked at the scale smaller than that concerned with observable CMB anisotropies $k\lesssim k_\lambda$.
Besides, the constraint Eq.~(\ref{CMBcons}) becomes looser for larger $k_c$, since the peak of the magnetic field power spectrum becomes apart from the CMB anisotropy scale.
Eq.~(\ref{CMBcons}) is more severe than Eq.~(\ref{COBEcons}) for $n<3.3$ and
$n<1.6$, when $k_c=10\,{\rm Mpc}^{-1}$ and $k_c=100\,{\rm Mpc}^{-1}$
respectively.

%%%%%%%%%%%%%%%%%%%%%%%%%%%%%%%%%%%%
%%%%%%%%%%%%%%%%%%%%%%%%%%%%%%%%%%%%

\subsection{Angular power spectra of CMB distortions}

Let us study angular power spectra of CMB
distortion anisotropies.  

\subsubsection{Correlations of CMB distortions}

\begin{figure}[tbp]
\begin{center}
\includegraphics[width=80mm]
{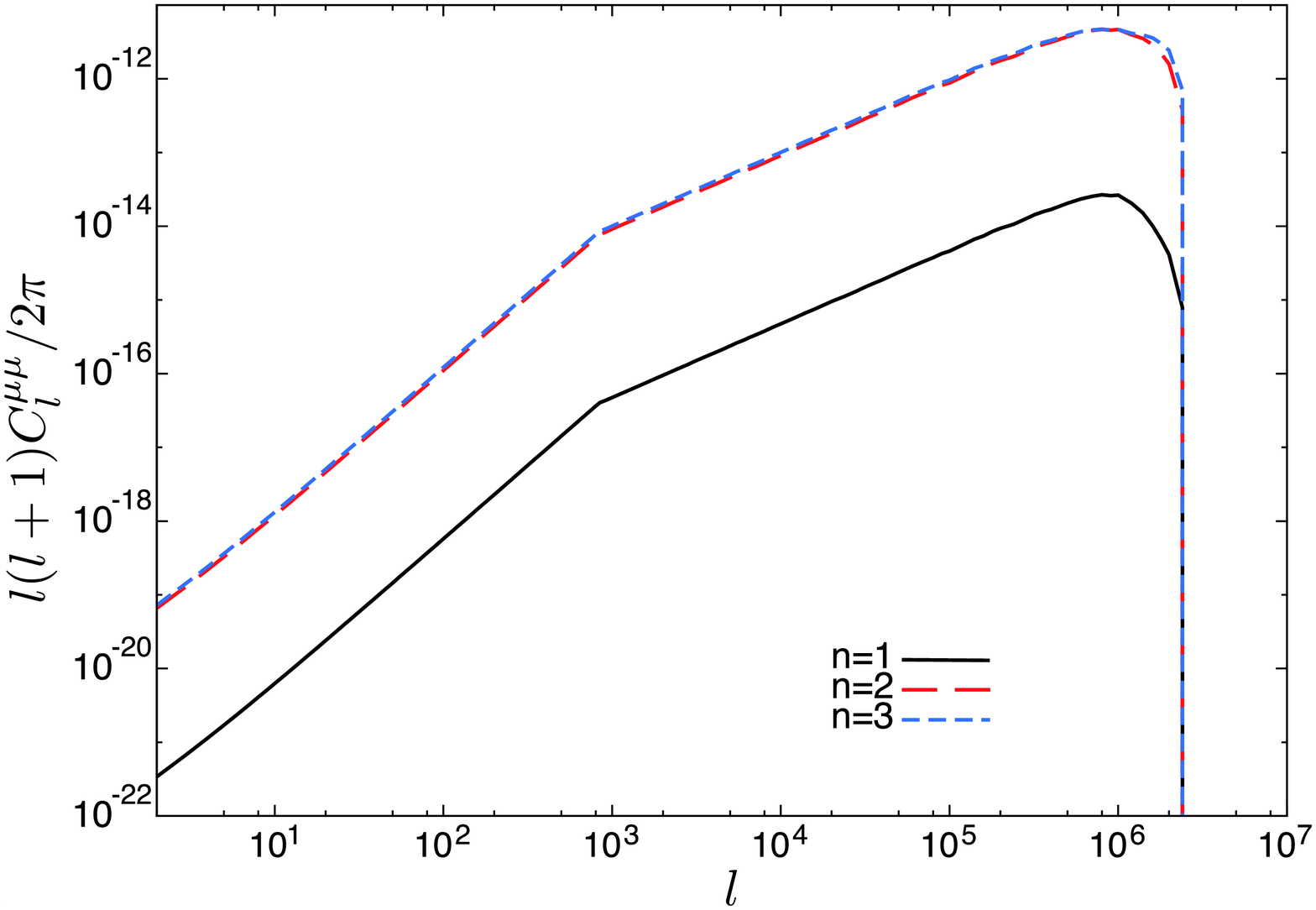}
\caption{The $\mu$-$\mu$ auto-correlation angular power spectra for $n=1$ (black solid), $n=2$ (red long dashed) and $n=3$(blue short dashed).
For all cases, $k_c=100~{\rm Mpc}^{-1}$.
$B_0$ is set to $B_0=32~{\rm nG}$ for $n=1$, which corresponds to
 $B_\lambda=3.0~{\rm nG}$, and $B_0=1.1\times 10^2~{\rm nG}$ for $n=2,3$.
}
\label{fig:C_mumu}

~\\
~\\

\includegraphics[width=80mm]
{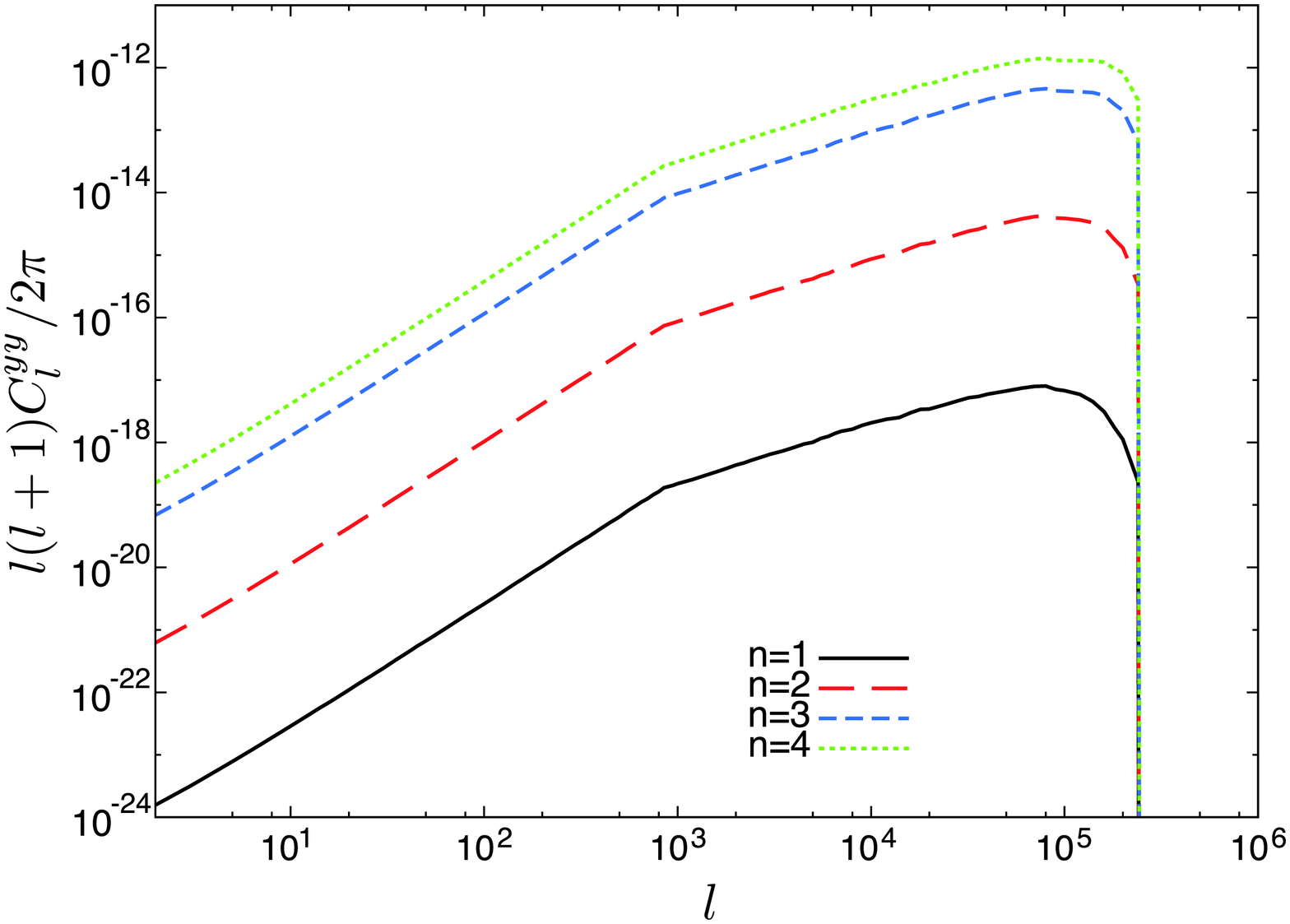}
\caption{The $y$-$y$ auto-correlation angular power spectra for $n=1$ (black solid), $n=2$ (red long dashed), $n=3$ (blue short dashed) and $n=4$ (green dotted).
For all cases, $k_c=10~{\rm Mpc}^{-1}$.
$B_0$ is set to $B_0=10~{\rm nG}$, $29~{\rm nG}$ and $83~{\rm nG}$ for $n=1$, 2 and 3 respectively, which corresponds to $B_\lambda=3.0~{\rm nG}$, and $B_0=1.1\times 10^2~{\rm nG}$ for $n=4$.
}
\label{fig:C_yy}

~\\
~\\

\includegraphics[width=80mm]
{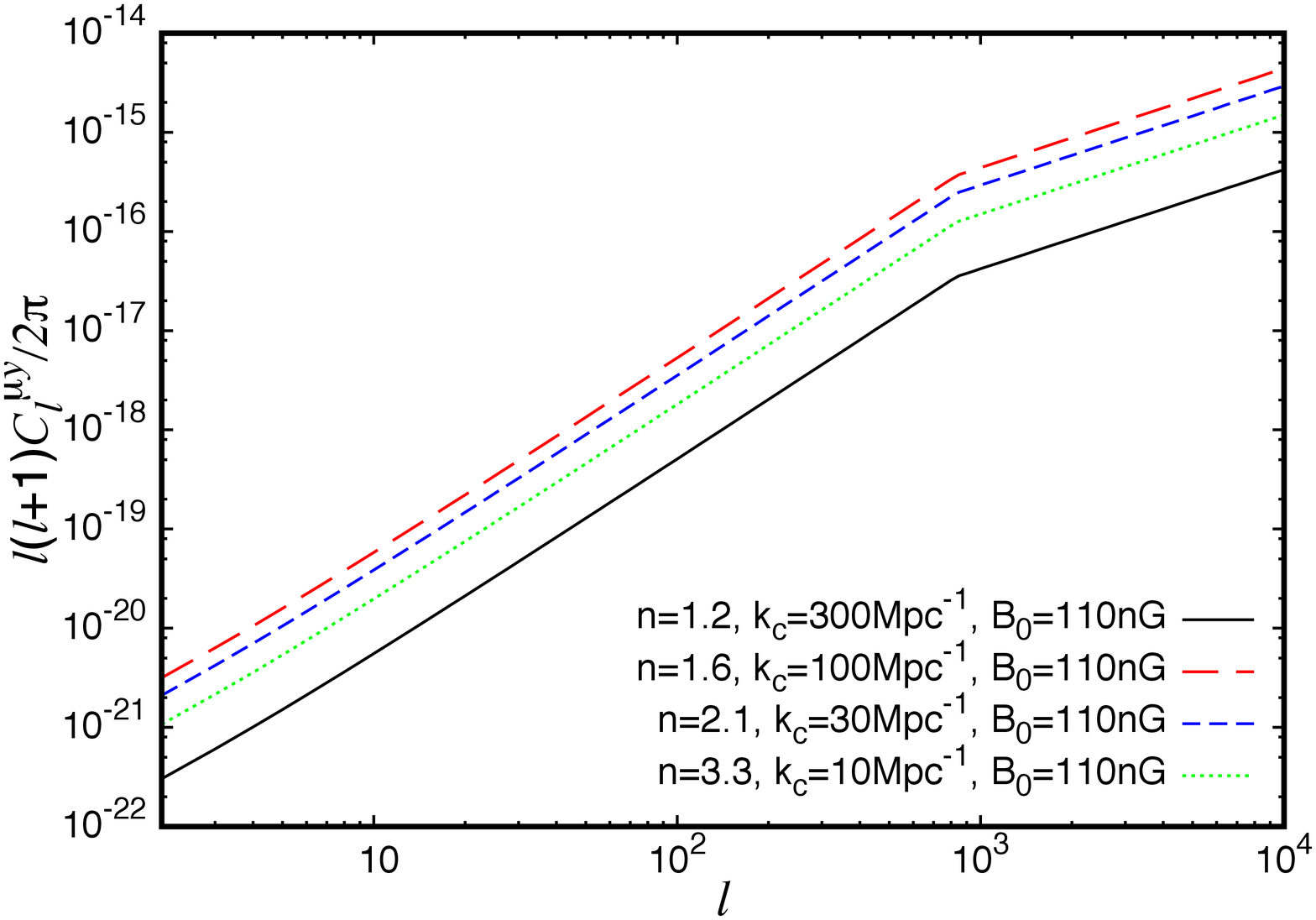}
\caption{The $\mu$-$y$ cross-correlation angular power spectra for parameter sets $(n, k_c[\rm{Mpc}^{-1}])=(1.2, 300)$ (black solid), 
$(1.6, 100)$ (red long dashed), $(2.1, 30)$ (blue short dashed) and $(3.3, 10)$ (green dotted).
For all cases, $B_0$ is fixed to $1.1\times 10^2~{\rm nG}$.
}
\label{fig:C_muy}
\end{center}
\end{figure}

First, we consider the auto- and cross-correlations in CMB distortions, i.e. $\mu$-$\mu$, $y$-$y$ and $\mu$-$y$.
Here, we choose the parameter sets so that we obtain the maximum amplitude of the power spectra
with the current constraint shown in FIG.~\ref{fig:n-B0} being satisfied. 
We show $C^{\mu\mu}_l$ in FIG.~\ref{fig:C_mumu} for $n=1$ (black solid), 2 (red long dashed) and 3 (blue short dashed),
where we fix $k_c=100~{\rm Mpc}^{-1}$.
Following FIG.~\ref{fig:n-B0},
we set $B_0$ to be a maximum allowed value for each $n$
as $32~{\rm nG}$ for $n=1$ and $1.1\times 10^2~{\rm nG}$ for $n=2$ and $n=3$.
We also show $C^{yy}_l$ in FIG. \ref{fig:C_yy} for $n=1$ (black solid), 2 (red long dashed), 3 (blue short dashed) and 4 (green dotted), 
where we fix $k_c=10~{\rm Mpc}^{-1}$.
As shown in FIG.~\ref{fig:n-B0},
for $k_c=10~{\rm Mpc}^{-1}$ the current observational cosmological limit for $B_0$ mainly comes from the CMB temperature anisotropies
(denoted as the red line) for $n \lesssim 3.5$ and it depends on the spectral index $n$.
Hence we set $B_0$ in FIG.~\ref{fig:n-B0} to be 
$10~{\rm nG}$, $29~{\rm nG}$, $83~{\rm nG}$ and $1.1\times 10^2~{\rm nG}$ for $n=1$, $n=2$, $n=3$ and $n=4$, respectively.
As for the cross angular power spectrum, $C^{\mu y}$, which is shown in FIG. \ref{fig:C_muy}, 
we fix the amplitude $B_0$ to be $1.1 \times 10^2$~nG and change the peak scale $k_c$ for each spectral index $n$.
Following the observational constraint from the CMB temperature anisotropies given by Eq.~(\ref{CMBcons})
and the relation between $B_\lambda$ and $B_0$ given by Eq.~(\ref{B0Blambda}),
for fixed $B_0$, the peak scale $k_c$ for each $n$ is chosen in order to obtain the maximum allowed value of $B_\lambda$. 
Then, in this figure, 
we set $k_c$ to 300, 100, 30, 10~Mpc$^{-1}$ for $n=1.2$ (black solid), 1.6 (red long dashed), 2.1 (blue short dashed) and 3.6 (green dotted), respectively.

 %%%%%%%%%%%%%%%%%%%%%%%%%%%%%%%%%%%%
%%%%%%%%%%%%%%%%%%%%%%%%%%%%%%%%%%%

The magnitudes of the auto-correlation spectra $C^{\mu\mu}_l$ and $C^{yy}_l$ can be roughly estimated as follows.
$C_X$ given by Eqs.~(\ref{Cmu}) or (\ref{Cy}) is
\be
C_X(k,k^\prime)\sim
\begin{cases}
1 & ;{\rm for} \ k_{X,f}<\max\{k,k^{\prime}\}<k_{X,i} \\
0 & ;{\rm otherwise}
\end{cases}, \label{CXkk}
\ee
where $k_{X,i}$ and $k_{X,f}$ 
the Fourier modes of magnetic fields which satisfies
$\tau (k_{X,i}, z_{X,i} ) = 1$
and
$\tau (k_{X,f}, z_{X,f} ) = 1$, respectively.
In this sense, 
$k_c$ is almost identical to $k_{X,f}$ here.
The spherical Bessel function can be approximated as
\be
j_l(x)\simeq 
\begin{cases}
0 & ;{\rm for} \ x<l \\
\frac{1}{x}\cos \left( x-\frac{(l+1)\pi}{2}\right) & ;{\rm for} \ x>l
\end{cases}.
\ee
Neglecting the effect of the finite thickness of the last scattering
surface, the auto-correlation spectrum $C^{XX}_l$ can be roughly estimated as
\be
\frac{l(l+1)C^{XX}_l}{2\pi}\sim \left(\frac{\rho_B}{\rho_{\gamma}}\right)^2\frac{l^2}{(k_{c}r_{\rm rec})^2}, \label{CXXest}
\ee
if $k_{X,f}<k_c<k_{X,i}$.
Since 
we are setting $k_c\sim k_{X,f}$ as mentioned before,
this estimation is consistent with the spectra shown in FIGs. \ref{fig:C_mumu}
and \ref{fig:C_yy}, especially with respect to the dependence on
$l$, $l(l+1)C^{XX}_l/2\pi\propto l^2$, for $l\lesssim 10^3$.
The reason why $l(l+1)C^{XX}_l/2\pi \propto l$ for
$l\gtrsim 10^3$ in FIGs. \ref{fig:C_mumu}
and \ref{fig:C_yy} is that the effect of the finite thickness of the last scattering surface, which is introduced as Eq. (\ref{finiteLS}),
suppresses $C^{XX}_l$ by a factor $l\sigma_{\rm LS}/r_{\rm rec}$.
The peak and the cut-off of $l(l+1)C^{XX}_l/2\pi$ at $l\sim 10^6$ for $\mu$ and at $l\sim 10^5$ for $y$
correspond to those of the magnetic fields power spectrum at $k=k_c$.

Eq. (\ref{CXXest}) shows that the amplitude of $C^{XX}_l$ is
determined by the total energy of decaying magnetic fields.  
Therefore, for fixed $k_c$, its amplitude is determined only by $B_0$, not by $n$.
Since, here, we  
set $B_0$ to a 
smaller value for $n=1$ than for $n=2$ and $n=3$ 
for $C^{\mu\mu}_l$
in order to satisfy the current observational constraints, the amplitude becomes also smaller.
Of course, $C^{\mu\mu}_l$ for $n=2$ and $n=3$ overlaps each other because of the same value of $B_0$ for both cases.
On the other hand, the amplitude of $C^{yy}_l$ apparently seems to depend on the spectral index $n$ in FIG. \ref{fig:C_yy}.
However, since for $C^{yy}_l$ we take smaller $k_c$ than that for $C^{\mu\mu}_l$,
the value of $B_0$ taken here strongly depends on the spectral index $n$
in order to maximize the amplitude within the observational constraints shown in FIG. \ref{fig:n-B0}, as we have shown above. 
In the range of
$n < 3.3$, $B_0$ is smaller for smaller $n$
and hence the amplitude of $C^{yy}_l$ becomes smaller for smaller $B_0$,
which is consistent with the simple estimation Eq.~(\ref{CXXest}).

The second factor of the RHS of Eq.~(\ref{CXXest}), which comes from the spherical Bessel function, 
tells us that smaller $k_c$ leads to larger $C^{XX}_l$ for fixed $l$, as mentioned in the previous subsection.
In fact, $r_{\rm rec} \simeq 10^4\, {\rm Mpc}^{-1}$ and we  
take $k_c=10~{\rm Mpc}^{-1}$ for $\mu$ and
$100~{\rm Mpc}^{-1}$ for $y$ here, and hence,
for the CMB observation scales ($l \lesssim 10^4$),
$C^{XX}_l$ is much suppressed compared with the value at the peak scale, $l \sim k_c r_{\rm rec}$.
This suppression reflects the fact that the typical scales of fluctuations of the
$\mu$- and $y$-parameter, $\sim 2\pi/k_c \sim 2\pi/k_{X,f}$ , are much smaller than the observation scale, $\sim r_{\rm rec}/l$.
Because we take smaller $k_c$ for $C^{yy}_l$ compared with $C^{\mu\mu}_l$, the amplitude of $C^{yy}_l$
seems to be larger than that of $C^{\mu\mu}_l$ for fixed $l$ and the same value of $B_0$.

On the other hand, as is shown in FIG.~\ref{fig:C_muy}, 
we found that the cross-correlation spectrum $C^{\mu y}_l$ in general cannot be as large as 
the auto-correlation ones $C^{\mu\mu}_l$ and $C^{yy}_l$ even if parameters $(n,~k_c,~B_0)$ are tuned. 
This is because the scales of primordial magnetic fields which mainly contribute to $\mu$- and $y$-type distortions
are different.
In other words, this can be understood by seeing that 
$C_\mu(\chi, q)C_y(\chi,q)$ in Eq.\eqref{CmuyB} vanishes, provided the rough approximation Eq. \eqref{CXkk}.
We thus conclude that it is difficult to observe $C^{\mu y}_l$ unless $C^{\mu\mu}_l$ and $C^{yy}_l$ 
are observed with high significance. Therefore we do not take into account $C^{\mu y}_l$ in Section \ref{sec:detect}, 
where we discuss detectability of primordial magnetic fields by CMB observations of distortion power spectra.

\subsubsection{Cross-correlation with CMB temperature anisotropies}

 %%%%%%%%%%%%%%%%%%%%%%%%%%%%%%%%%%%%
\begin{figure}[htbp]
\begin{center}
\includegraphics[width=80mm]
{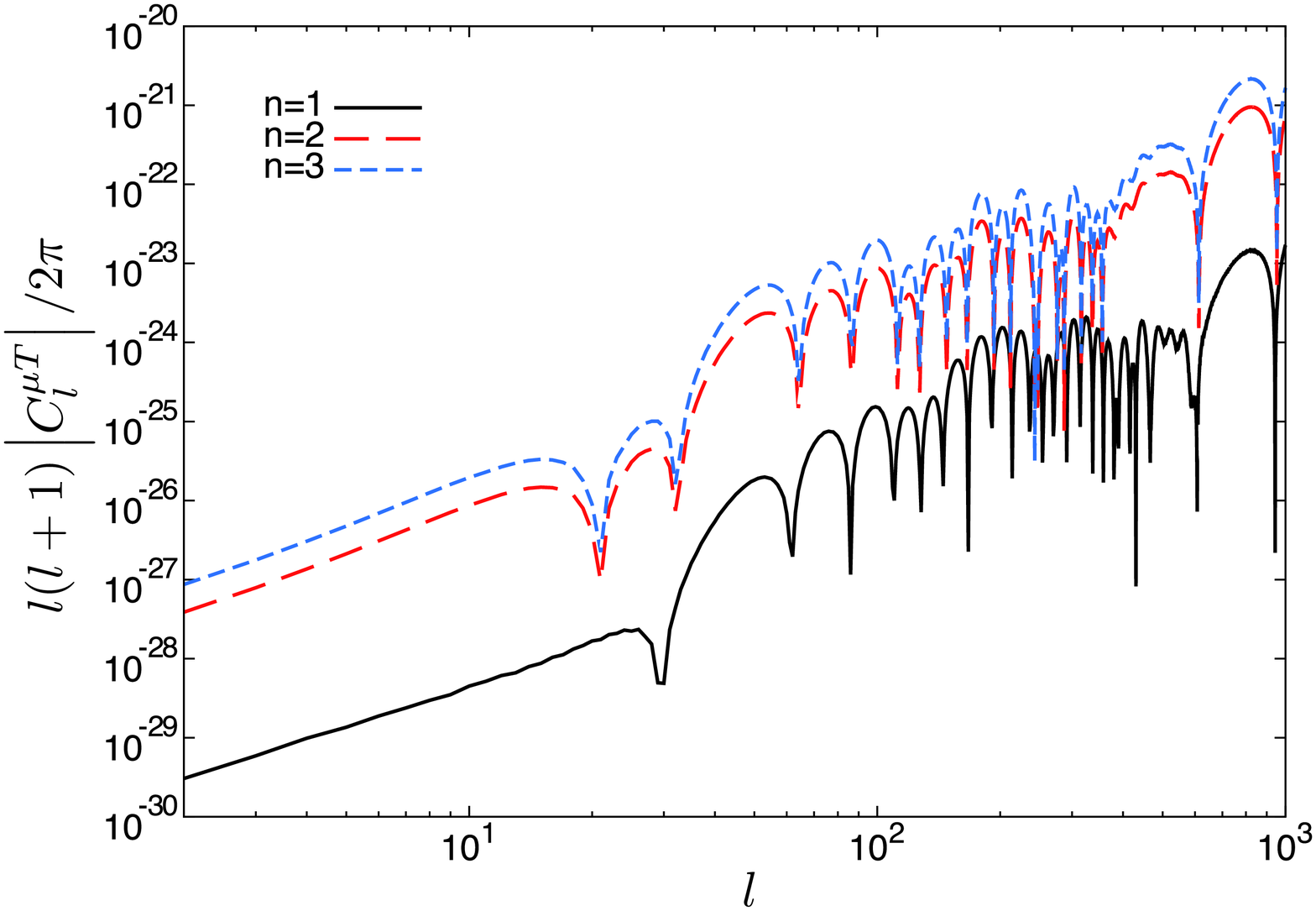}
\caption{The $\mu$-$T$ cross-correlation angular power spectra for 
$n=1$ (black solid), $n=2$ (red long dashed) and $n=3$ (blue short dashed).
For all cases, $k_c=100~{\rm Mpc}^{-1}$.
$B_0$ is set to $B_0=32~{\rm nG}$ for $n=1$,
and $B_0=1.1\times 10^2~{\rm nG}$ for $n=2,3$.}
\label{fig:C_muT}

~\\
~\\

\includegraphics[width=80mm]
{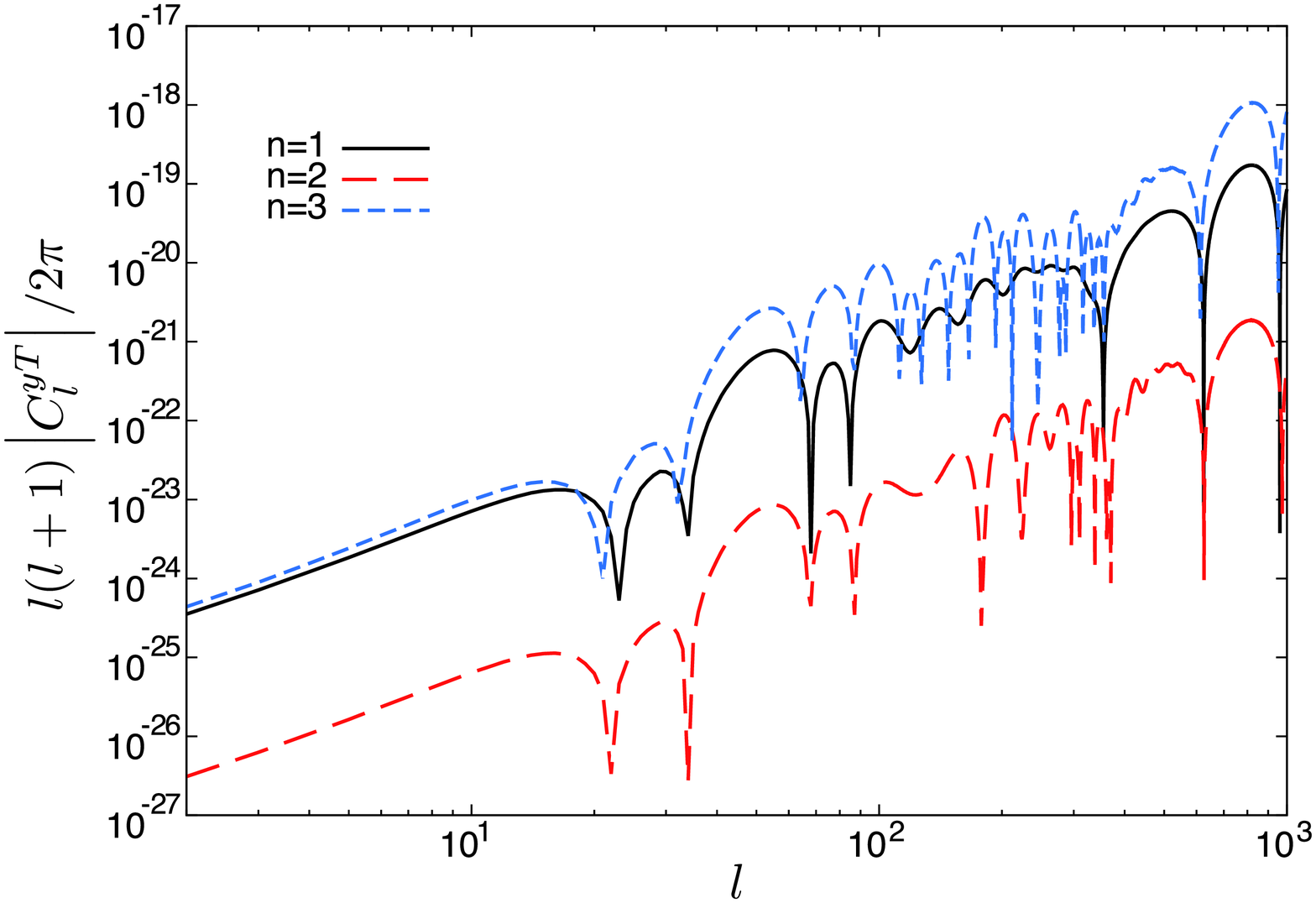}
\end{center}
\caption{The $y$-$T$ cross-correlation angular power spectra for 
$n=1$ (black solid), $n=2$ (red long dashed) and $n=3$ (blue short dashed).
For all cases, $k_c=10~{\rm Mpc}^{-1}$.
$B_0$ is set to $B_0=10~{\rm nG}$, $29~{\rm nG}$ and $83~{\rm nG}$ for $n=1$, 2 and 3 respectively, which corresponds to $B_\lambda=3.0~{\rm nG}$.}
\label{fig:C_yT}
\end{figure}

Next, we calculate the cross-correlation between CMB distortion anisotropies and the scalar passive mode of CMB temperature anisotropies
given by Eqs. (\ref{CmuTB}) and (\ref{CyTB}).
We show $C^{\mu T}_l$ in FIG. \ref{fig:C_muT} for
$n=1$ (black), 2 (red long dashed) and 3 (blue short dashed).
For all cases, we set $k_c=100~{\rm Mpc}^{-1}$
and $B_0=32 {\rm nG}$ for $n=1$, which corresponds to $B_\lambda=3.0~{\rm nG}$, and $B_0=1.1\times 10^2~{\rm nG}$ for $n=2,3$ respectively.
We also show $C^{y T}_l$ in FIG. \ref{fig:C_yT} for 
$n=1$ (black), 2 (red long dashed) and 3 (blue short dashed).
We take $k_c=10~{\rm Mpc}^{-1}$ 
and $B_0=10~{\rm nG}$, $29~{\rm nG}$ and $83~{\rm nG}$ for $n=1$, 2 and 3, respectively.
As shown in
 FIGs. \ref{fig:C_muT} and \ref{fig:C_yT}, $C^{\mu T}_l$ and $C^{y
T}_l$ are suppressed compared with the auto-correlations $C^{\mu \mu}_l$ and $C^{yy}_l$.  This
is just because the typical length scale of the CMB distortion fluctuations is
much smaller than the Silk damping scale $k_{\rm Silk}\sim 0.1~{\rm Mpc}^{-1}$.
As is well known, the CMB temperature fluctuations have been exponentially
damped by the Silk damping~\cite{Silkdamping}
on the scales with $k > k_{\rm Silk}$.
On the other hand, 
the amplitudes of the anisotropies of $\mu$ and $y$
on the CMB observation scales ($k < k_{\rm Silk}$), where the CMB temperature anisotropy
keeps its amplitude,
are much suppressed as shown in the above discussion about the auto power spectra of $\mu$ and $y$.
Hence, even though the cross correlations between the temperature and distortion anisotropies
exist due to the fact that both of these anisotropies are given in terms of the convolution of the Gaussian magnetic fields,
the cross correlations are more suppressed than auto correlations of those anisotropies.
Note that as shown in FIGs. \ref{fig:C_muT} and \ref{fig:C_yT},
the amplitudes of the cross-correlations depend on not only $B_0$ but also $n$ in contrast with the case
of auto power spectra of $\mu$ and $y$.  
This is because the $\mu$ and $y$ anisotropies on the scales larger than the Silk scale ($k < k_{\rm Silk}$)
depend on not only $B_0$, which determines the amplitudes of $\mu$ and $y$ anisotropies on the peak scale $\sim 2\pi/k_c$, but also the tilt $n$.

\section{Detectability of CMB distortion anisotropies} \label{sec:detect}

In this section, we study the detectability of the anisotropies of CMB
distortion parameters in the future observations by performing a signal-to-noise (SN) analysis.
In order to evaluate SN ratio (SNR), first, we must estimate the variance of the
angular power spectrum.
The variances of $C^{\mu\mu}_l$ and $C^{yy}_l$ estimated from the full sky observation of CMB are given by \cite{Knox:1995dq}
\be
\sigma^2_{ll^\prime}=\left< \left(C^{XX}_l-\left<C^{XX}_l\right>\right) \left(C^{XX}_{l^\prime}-\left<C^{XX}_{l^\prime}\right>\right)\right>=\frac{2}{2l+1}\left(C^{XX}_l+C^{XX,N}_l\right)^2\delta_{ll^\prime},
\ee
where  
$C^{XX,N}_l$ is the noise power
spectrum of the observation. We assume that the foregrounds can be
removed perfectly. In this assumption, the noise power spectrum $C^{XX,N}_l$ consists
of the experimental noise power spectrum and 
can be written as \cite{Knox:1995dq}
\be
C^{XX,N}_l=\sigma_X^2\theta_b^2 b_l^{-2},
\label{eq:auto_cov}
\ee
where $\sigma_X$ is the $1\sigma$ uncertainty in $X$ per pixel, $\theta_b$ is the beam width and $b_l$ is the so-called beam transfer function given by
\be
b_l=\exp \left(-\frac{l^2\theta_b^2 }{16\ln 2}\right).
\ee
From Eq.~(\ref{eq:auto_cov}),
we can obtain the SNR in a measurement of $C^{\mu\mu}_l$ and $C^{yy}_l$ by
\be
\left(\frac{S}{N}\right)^2=\sum_l\frac{2l+1}{2}\frac{\left(C^{XX}_l\right)^2}{\left(C^{XX}_l+C^{XX,N}_l\right)^2}.
\ee
We define 
a function which represents 
the detectable level of the signal $l(l+1)C^{XX}_l/2\pi$, $C_l^{DL}$, as
\be
C_l^{DL}=\frac{l(l+1)}{2\pi}\sqrt{\frac{2}{(2l+1)l}}C^{XX,N}_l. \label{effnoise}
\ee
The fact that 
$l(l+1)C^{XX}_l/2\pi>C_l^{DL}$
means that SNR becomes larger than 1. 

When we take a logarithmically homogeneous binning of $l$ with bin width $\Delta \ln l=1$, there are $l$ multipoles in a bin at $l$.
Since different multipoles are independent, the noise level per each bin should be given by $\sigma_{ll}/\sqrt{l}$.
Therefore, the detectable level of $l(l+1)C^{XX}_l/2\pi$ is roughly given by Eq. (\ref{effnoise}).

For the cross-correlations, $C^{\mu T}_l$ and $C^{yT}_l$, the variance
is obtained from
\be
\sigma^2_{ll^\prime}=\left< \left(C^{XT}_l-\left<C^{XT}_l\right>\right) \left(C^{XT}_{l^\prime}-\left<C^{XT}_{l^\prime}\right>\right)\right>=\frac{1}{2l+1}\left(C^{XX}_l+C^{XX,N}_l\right)\left(C^{TT}_l+C^{TT,N}_l\right)\delta_{ll^\prime},
\ee
where $C^{TT}_l$ is the primary CMB temperature power spectrum
and $C^{TT,N}_l$ is the noise
power spectrum for the CMB temperature observation.
Here
we assume that, compared with the CMB signal, the
experimental noise is very small on scales of interest. Therefore, we can
neglect the noise power spectrum.
In this assumption, the SNR for the cross-correlations is given by
\be
\left(\frac{S}{N}\right)^2\simeq\sum_l(2l+1)\frac{\left(C^{XT}_l\right)^2}{\left(C^{XX}_l+C^{XX,N}_l\right)C^{TT}_l}.
\ee

Let us discuss the detectability of anisotropic CMB distortions in each future experiment.

\subsection{PIXIE case}

\begin{figure}[t]
\begin{tabular}{cc}
\begin{minipage}{0.5\hsize}
\begin{center}
  \subfigure[The detectable level of $l(l+1)C^{\mu\mu}_l/2\pi$.]{
\includegraphics[width=75mm]
{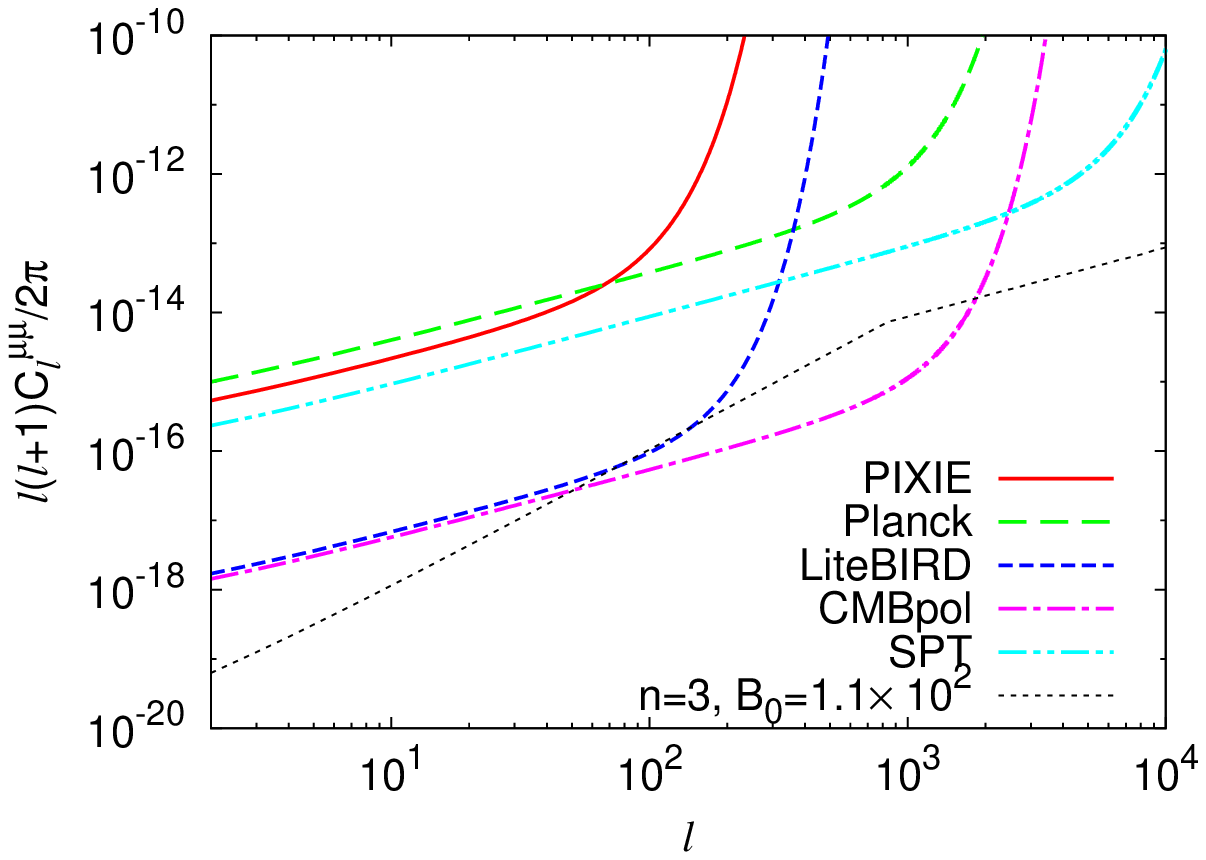}
\label{fig:CmuN}
}
\end{center}
\end{minipage}

\begin{minipage}{0.5\hsize}
\begin{center}
  \subfigure[The detectable level of $l(l+1)C^{yy}_l/2\pi$.]{
\includegraphics[width=75mm]
{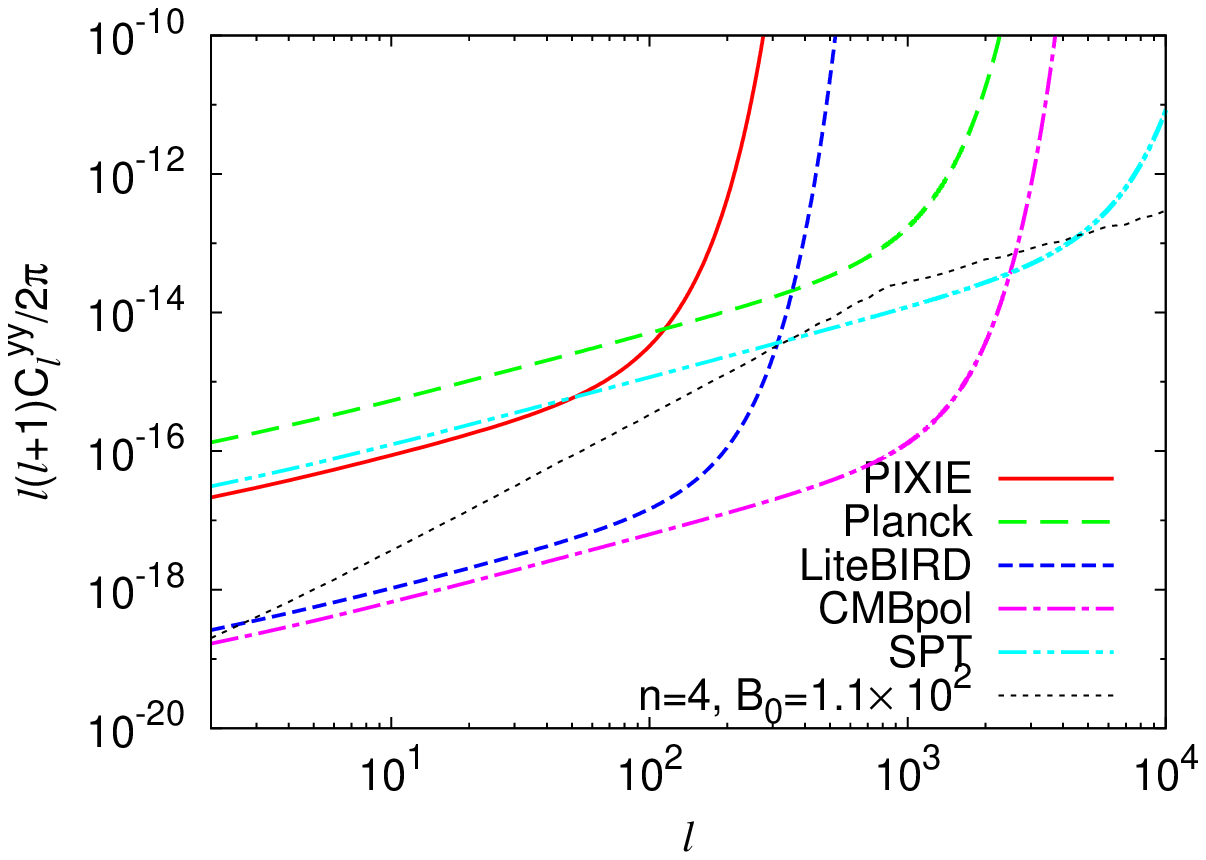}
\label{fig:CyN}
}
\end{center}
\end{minipage}
\end{tabular}
\caption{The detectable levels of $l(l+1)C^{\mu\mu}_l/2\pi$ and $l(l+1)C^{yy}_l/2\pi$ in PIXIE (red solid), Planck (green long dashed), LiteBIRD (blue short dashed), CMBpol (magenta chain) and SPT (light blue two-dot chain).
We also plot $l(l+1)C^{\mu\mu}_l/2\pi$ for $n=3, B_0=1.1\times 10^2~{\rm
 nG}$ and $l(l+1)C^{yy}_l/2\pi$ for $n=4, B_0=1.1\times 10^2~{\rm nG}$
 for comparison (black dotted).}
\label{fig:CN}
\end{figure}

PIXIE \cite{Kogut:2011xw} is a recently proposed satellite for CMB observation, which can measure CMB distortion parameters to 
a very high accuracy.
For PIXIE, the beam width is $\theta_b=1.6^{\circ}$ and the $1\sigma$ uncertainty in $\mu$ and $y$ parameters averaged over the full sky are $\delta \mu=10^{-8}$ and $\delta y =2\times 10^{-9}$ respectively~\cite{Kogut:2011xw}.
This leads to
\be
C^{\mu\mu,N}_l=1.3\times 10^{-15}\times \exp
\left(\frac{l^2}{84^2}\right) , \label{CmumuNPIXIE}
\ee
and
\be
C^{yy,N}_l=5.0\times 10^{-17}\times \exp \left(\frac{l^2}{84^2}\right).\label{CyyNPIXIE}
\ee
In FIG. \ref{fig:CN}, we show the function of the detectable level, $C_l^{DL}$, of 
the auto power spectra of $\mu$ (left panel (a)) and $y$ (right panel (b)) for each CMB experiment
and 
the largest $C^{XX}_l$ allowed by the current observations with a black dotted line.
The detectable level function for PIXIE
is shown
as a function of the multipole $l$ with a red solid line.
Due to the exponential factor in Eqs.~(\ref{CmumuNPIXIE}) and (\ref{CyyNPIXIE}), PIXIE can measure only large scale anisotropies, which
correspond to $l\lesssim 100$, and for both of $\mu$ and $y$ distortions
the black dashed line, which represents the largest signal allowed by the current observations,
is much lower than the red line. 
As a result, it would be difficult to detect
the auto-correlation signals of
the CMB distortion anisotropies induced by primordial magnetic
fields by PIXIE.

The cross-correlation signals between CMB distortion and temperature anisotropies
would be more difficult to be detected by PIXIE.
A rough estimate gives $l(l+1)C^{TT}_l\sim 6.0\times 10^{-10}$ as in \cite{Pajer:2012vz}, and
we see that it is necessary that $C^{\mu T}_{l=100} \gtrsim 10^{-16}$ or $C^{y T}_{l=100} \gtrsim 10^{-17}$ for the SNR larger than 1.
Both of $C^{\mu T}_l$ and $C^{yT}_l$ shown in FIGs.~\ref{fig:C_muT} and \ref{fig:C_yT} are much smaller than these required values.

\subsection{Planck case}

The authors of \cite{Ganc:2012ae} have argued that the anisotropies of CMB
distortion parameters can be detected not only absolutely calibrated
experiments such as PIXIE but also relatively calibrated experiments
like WMAP and Planck, although an isotropic CMB distortion can be probed
only by absolutely calibrated experiments.
For relatively calibrated experiments, anisotropies of CMB
distortion parameters are seen as temperature anisotropies, whose
amplitude depends on the frequency channel.  The temperature for photon
frequency $\nu$ is given by
\be
T(\nu)=\frac{T_0x}{\ln (1+n(x))^{-1}},
\ee
where $x=2\pi\nu/T_0$, $T_0$ is the temperature averaged over all sky
and $n(x)$ is the photon occupation number.
Without CMB distortions, the occupation number is given by 
the Planck distribution as $n(x)=(e^x-1)^{-1}\equiv n_0(x)$.
Due to the CMB distortions the energy spectrum of CMB photons deviates
from the Planck distribution. As a result, 
the apparent temperature anisotropy depending on the frequency channel
is created from the CMB distortions.
In the case of the $\mu$-distortions,  the apparent temperature
anisotropy is given by~\cite{Ganc:2012ae}
\be
\frac{\delta T(\hat n, \nu)}{T} \simeq -\frac{\delta \mu(\hat n)}{x},
\ee
where $\delta \mu$ is the fluctuating part of $\mu$ and $x=2\pi \nu/T$.
In the case of $y$-distortion, the resultant temperature anisotropy is
\be
\frac{\delta T(\hat n, \nu)}{T}=\left(x\frac{e^x+1}{e^x-1}-4\right)\delta y(\hat n)\equiv a(\nu)\delta y(\hat n),
\ee
where $\delta y$ is the fluctuating part of $y$.
As shown in the above expressions, 
these temperature anisotropies produced by the CMB distortions have the
frequency dependence, and taking difference between the
temperature anisotropies in different frequency channels,
we find  $\delta \mu$ and $\delta y$.
The experimental noise power spectrum in this type of observation using
two different frequency channels $\nu_1$ and $\nu_2$ is given by
\be
C^{\mu\mu,N}_l=\left[\frac{\nu_1\nu_2/(\nu_1-\nu_2)}{56.80{\rm GHz}}\right]^2\sum_{i=1,2}\sigma^2_{T,i}\theta_{b,i}^2b^{-2}_{i,l},
\ee
for $\mu$-distortions and
\be
C^{yy,N}_l=\left(\frac{1}{a(\nu_1)-a(\nu_2)}\right)^2\sum_{i=1,2}\sigma^2_{T,i}\theta_{b,i}^2b^{-2}_{i,l},
\ee
for $y$-distortions, where $\sigma_{T,i}$ is the $1\sigma$ uncertainty in $\delta T/T$ per pixel at frequency $\nu_i$, $\theta_{b,i}$ is the beam width of channel $\nu_i$ and $b_{i,l}=\exp \left(-l^2\theta_{b,i}^2 /16\ln 2\right)$.

\begin{table}[t]
\begin{center}
\begin{tabular}{cc}
\begin{minipage}{0.3\hsize}
\begin{center}
\subtable[Planck (from \cite{Planck:2006aa})]{
  \begin{tabular}{c|c|c}
  \hline
  \hline
  bands [GHz] & $\theta_b$  & $\sigma_T$ \\
  \hline
      $100$ & $9.5^\prime$ & $2.5\times 10^{-6}$ \\
      $143$ & $7.1^\prime$ & $2.2\times 10^{-6}$  \\
  \hline
  \hline 
\end{tabular}
\label{table:planck}
}
\end{center}
\end{minipage}

\begin{minipage}{0.3\hsize}
\begin{center}
\subtable[LiteBIRD (from \cite{LiteBIRD})]{
  \begin{tabular}{c|c|c}
  \hline
  \hline
  bands [GHz] & $\theta_b$  & $\sigma_T$ \\
  \hline
      $90$ & $60^\prime$ & $2.1\times 10^{-8}$ \\
      $150$ & $36^\prime$ & $3.3\times 10^{-8}$  \\
  \hline
  \hline 
\end{tabular}
\label{table:LiteBIRD}
}
\end{center}
\end{minipage}
\\
\begin{minipage}{0.3\hsize}
\begin{center}
\subtable[CMBpol (from \cite{Baumann:2008aq})]{
  \begin{tabular}{c|c|c}
  \hline
  \hline
  bands [GHz] & $\theta_b$ & $\sigma_T$ \\
  \hline
      $100$ & $8^\prime$ & $1.1\times 10^{-7}$ \\
      $150$ & $5^\prime$ & $1.6\times 10^{-7}$  \\
  \hline
  \hline 
\end{tabular}
\label{table:CMBpol}
}
\end{center}
\end{minipage}

\begin{minipage}{0.3\hsize}
\begin{center}
\subtable[SPT (from \cite{Reichardt:2011yv,Schaffer:2011mz})]{
  \begin{tabular}{c|c|c}
  \hline
  \hline
  bands [GHz] & $\theta_b$ & $\sigma_T$ \\
  \hline
      $95$ & $1.7^\prime$ & $9.6\times 10^{-6}$ \\
      $150$ & $1.2^\prime$ & $5.5\times 10^{-6}$  \\
  \hline
  \hline 
\end{tabular}
\label{table:SPT}
}
\end{center}
\end{minipage}

  \end{tabular}
\addtocounter{table}{-1}
  \caption{Parameters which characterize sensitivities of various relatively calibrated experiments.
  $\theta_b$ is Gaussian beam width at FWHM, 
  $\sigma_T$ is temperature noise.
  Note that each satellite has some frequency bands other than those shown above.
  We here show parameters for two bands which have best sensitivities.
  }

\end{center}
\end{table}
\stepcounter{table}

According to \cite{Planck:2006aa}, 
Planck has the sensitivity $\sigma_T=2.5\times 10^{-6}$ with the beam
width $\theta_b=9.5^\prime$ for $100~{\rm GHz}$ channel
and $\sigma_T=2.2\times 10^{-6}$ with $\theta_b=7.1^\prime$
for $143~{\rm GHz}$ shown in TABLE \ref{table:planck}.
These channels have the best sensitivity among frequency channels of Planck.
Therefore, the noise power spectrum for $\mu$ and $y$-distortions are 
\be
C^{\mu\mu,N}_l=1.6\times 10^{-15}\times e^{(l/855)^2}+7.1\times 10^{-16}\times e^{(l/1.1\times 10^3)^2},
\ee
and
\be
C^{yy,N}_l= 2.2\times 10^{-16}\times e^{(l/855)^2}+9.4\times 10^{-17}\times e^{(l/1.1\times 10^3)^2},
\ee
respectively.
From the above expressions, we find that
Planck is expected to probe the anisotropies on smaller scales compared with PIXIE due to the difference of exponential factor.
However, as shown in FIG.~\ref{fig:CN}
where $C_l^{DL}$ for Planck is shown as a green long-dashed line,
both of the largest allowed $C^{\mu\mu}_l$ and $C^{yy}_l$
still do not have amplitudes large enough to be detected by Planck.
Actually, $C^{\mu\mu}_l$ for $n=3, B_0=1.1\times 10^{2}{\rm nG}$ gives
$S/N\simeq 1.3\times 10^{-2}$ and $C^{yy}_l$ for $n=4,
B_0=1.1\times 10^{2}{\rm nG}$ gives $S/N\simeq 0.3$.
For the cross-correlation cases,
$C^{\mu T}_l$ and $C^{yT}_l$ are far from the detectable level: $C^{\mu T}_l \gtrsim 10^{-18}$ and $C^{y T}_l \gtrsim 10^{-19}$ at $l\sim 10^3$.

\subsection{LiteBIRD case}

LiteBIRD~\cite{LiteBIRD} is a proposed CMB satellite which aims to detect low $l$ B-mode polarization anisotropy.
Although the angular resolution of LiteBIRD will be worse than Planck,
the sensitivity per pixel will be better than Planck
and it would be a powerful experiment to detect the CMB distortions through the relative calibration.
For the beam width and sensitivity for the $90~{\rm GHz}$ and the $150~{\rm GHz}$ channels
shown in TABLE \ref{table:LiteBIRD}, the experimental noise power spectrum for the relative calibration by LiteBIRD
is given by
\be
C^{\mu\mu,N}_l=2.2\times 10^{-18}\times e^{(l/135)^2}+1.8\times 10^{-18}\times e^{(l/226)^2},
\ee
and
\be
C^{yy,N}_l= 3.3\times 10^{-19}\times e^{(l/135)^2}+2.8\times 10^{-19}\times e^{(l/226)^2}.
\ee
From the above expressions, we find that LiteBIRD can probe $\mu$ and $y$-distortion anisotropies with a
sensitivity better than PIXIE up to as high $l$ as PIXIE can.
As a result, 
in an
observation by LiteBIRD, anisotropies of CMB distortions induced by primordial
magnetic fields 
can reach the detectable level
within satisfying the current observational constraints,
as shown in Fig.~\ref{fig:CN} where
$C_l^{DL}$ for LiteBIRD is shown as a blue short-dashed line.
In particular, $C^{yy}_l$ for
$n=3, B_0=1.1\times 10^{2}~{\rm nG}$ gives $S/N\simeq 8$ and
that for $n=4, B_0=1.1\times 10^{2}~{\rm nG}$ gives $S/N\simeq 22$.
Approximating that $C^{yy}_l$ depends on $B_0$ only through the
overall factor proportional to $B_0^4$ \footnote{ Strictly speaking,
$B_0$ affects $C^{yy}_l$ also through ${\rm Im}\, \omega(t,k)$
in Eq. (\ref{Imomega}).  } and setting the detection threshold of the SNR to 4,
we find the threshold value of $B_0$ for detection of $C^{yy}_l$.
For $n\gtrsim3$, magnetic fields with
$B_0>70~{\rm nG}$ and 
$k_c = 10{\rm Mpc}^{-1}$ can generate detectable anisotropies of $y$-distortions.
For a smaller value of $n$, $C^{yy}_l$ can not be observable with the constraint Eq.~(\ref{CMBcons}) satisfied.
$C^{\mu\mu}_l$ is slightly smaller than the detectable level ($C^{\mu\mu}_l$ for $n=3, B_0=1.1\times 10^{2}{\rm nG}$ gives $S/N\simeq 1.3$)
and the cross correlations,
$C^{\mu T}_l$ and $C^{yT}_l$, are still far too small to be detected by LiteBIRD.

\subsection{CMBpol case}

CMBpol~\cite{Baumann:2008aq} is a future CMB satellite with a sensitivity similar to LiteBIRD and an angular resolution higher than Planck.
Using the $100{\rm GHz}$ and $150{\rm GHz}$ bands, whose beam width and
sensitivity are given in TABLE \ref{table:CMBpol},
we evaluate each noise power spectrum of $\mu$ and $y$ parameter as
\be
C^{\mu\mu,N}_l=1.6\times 10^{-18}\times e^{(l/1.0\times10^3)^2}+1.6\times 10^{-18}\times e^{(l/1.6\times10^3)^2},
\ee
and
\be
C^{yy,N}_l= 1.9\times 10^{-19}\times e^{(l/1.0\times10^3)^2}+1.8\times 10^{-19}\times e^{(l/1.6\times10^3)^2}.
\ee
Because of a high sensitivity and a high angular resolution, CMBpol enlarges the possibility to probe anisotropies of $y$-distortions,
as shown in Fig.~\ref{fig:CN} where
$C_l^{DL}$ for CMBpol is shown as a magenta dot-dashed line.
Setting the detection threshold of the SNR to 4 in a similar way to the LiteBIRD case,
we find that the auto power spectrum of $y$ anisotropies
can be detected for the primordial magnetic fields with 
$n\gtrsim 3$, $B_0>39~{\rm nG}$ and $k_c=10~{\rm Mpc}^{-1}$.
On the other hand, if the power spectrum of primordial magnetic fields
is less blue-tilted, that is, $n\lesssim 2$,
the primordial magnetic fields satisfying the current constraint Eq.~(\ref{CMBcons}) cannot induce detectable anisotropies of $y$.
In CMBpol experiment, $\mu$ anisotropies can also reach the detectable level as shown in FIG.~\ref{fig:CmuN}.
For $n\gtrsim 2$, in particular, the primordial
magnetic fields with $B_0>60~{\rm nG}$ and $k_c=100~{\rm Mpc}^{-1}$ create $C^{\mu\mu}_l$ detectable with a SNR larger than 4.
However the smaller value of $n$ makes $\mu$ anisotropies undetectable
as in the case of $y$ distortions.
For the cross-correlation signals,
$C^{\mu T}_l$ and $C^{yT}_l$ still can not be detected by CMBpol.

\subsection{SPT case}

Angular power spectra of $\mu$ and $y$ parameters induced by primordial
magnetic fields have a larger amplitude on small scales as mentioned in the previous section, where
space-based CMB experiments such as the above examples can not probe.  CMB
observations on such small scales $l> 1000$ are performed by
ground-based telescopes.
South Pole Telescope (SPT)~\cite{Reichardt:2011yv} is one of the latest ones among such telescopes.
According to \cite{Reichardt:2011yv,Schaffer:2011mz}, 
SPT has $\sigma_T=9.6\times 10^{-6}$ and
$\theta_b=1.7^{\prime}$ for the
$95~{\rm GHz}$ band and
$\sigma_T=5.5\times 10^{-6} $ and $ \theta_b=1.2^{\prime}$ for the $150~{\rm GHz}$ band shown in TABLE \ref{table:SPT}.
The noise power spectrum of SPT for each type of distortion is respectively given by
\be
C^{\mu\mu,N}_l=4.2\times 10^{-16}\times
e^{(l/4.8\times10^3)^2}+7.6\times 10^{-17}\times
e^{(l/6.8\times10^3)^2}, \ee
and 
\be
C^{yy,N}_l=1.4\times 10^{-17}\times
e^{(l/4.8\times10^3)^2}+1.0\times 10^{-17}\times
e^{(l/6.8\times10^3)^2}. \ee
The detectable level, $C_l^{DL}$, for SPT is shown as a light blue two-dot chain line in FIG. \ref{fig:CN}.
From FIG. \ref{fig:CyN}, we can see that SPT can detect $C^{yy}_l$ induced by the primordial magnetic fields, while as we will discuss shortly later
contamination from the SZ effect would be significant. On the other hand, from FIG. \ref{fig:CmuN}, $C^{\mu\mu}_l$ induced from 
primordial magnetic fields satisfying the current observational constraints is too small to be detected by SPT.

\subsection{Effects of the thermal Sunyaev-Zel'dovich effect}

So far we have discussed the detectability of $\mu$- and $y$-distortion anisotropies induced by 
primordial magnetic fields, simply assuming that there are no other sources of the distortions.
However, these distortions can be generated by various processes both in the early and late-time Universe.
In particular, as is mentioned in the introduction, $y$-distortion is generated by the thermal SZ effect in the late-time Universe. 
According to a recent measurement of $y$-distortion map 
by the Planck satellite \cite{Ade:2013qta}, $C^{yy}_l$ generated by the SZ effect  is $\mathcal O(10^{-16})$
at $50\lesssim\ell\lesssim1000$. This is about three order of magnitude larger than the maximum 
$C^{yy}_l$ from primordial magnetic fields (See Fig. \ref{fig:CyN}). Since, at large angular scales, both $C^{yy}_l$ from
the thermal SZ effect and primordial magnetic fields have the same spectral shapes $C^{yy}_l\sim\rm{constant}$, 
it should be difficult to detect the $C^{yy}_l$ from primordial magnetic fields. 
However, at smaller scales, shapes of the spectra differ. In particular, $C^{yy}_l$ of the thermal SZ effect drops sharply
at around the angular scales of galaxy clusters, $l\sim 3000$, while one from primordial magnetic fields 
decays mildly as $C^{yy}_l\propto 1/l$. Thus, we expect future observations which observe 
$C^{yy}_l$ at small angular scales may be able to distinguish primordial magnetic fields and the SZ effect.
Furthermore, we also expect that the cross-correlation between thermal SZ effect and and the distribution of galaxy clusters (see, e.g., Ref.~\cite{Fang:2011zk})
helps us distinguish between the signals of $y$-distortion from the thermal SZ effect and the primordial magnetic fields considered here.
On the other hand, since $\mu$-distortions cannot be generated by astrophysical processes, 
we expect that it is more plausible to seek for the signature of primordial magnetic fields 
in the $\mu$-distortion.

\section{Conclusion}

In this paper, we have considered $\mu$- and $y$-distortions of the
CMB photon energy spectrum, which are generated by decay of
primordial magnetic fields.  In particular, we have focused on
anisotropies of the CMB distortions, which are induced by space-varying
random magnetic fields.  Using the decay rate of magnetic fields derived
in \cite{Jedamzik:1996wp}, we have presented the formalism to calculate
the angular power spectra of these distortion parameters.
We have also considered the cross-correlations between the CMB distortion parameters and
temperature anisotropies induced by magnetic fields.

We have numerically calculated angular power
spectra $C^{\mu\mu}_l$, $C^{yy}_l$, {$C^{\mu y}_l$,} $C^{\mu T}_l$ and $C^{yT}_l$, taking
various values of the tilt of the magnetic field power spectrum $n$.
We have evaluated the maximum values 
of $C^{\mu\mu}_l$, $C^{yy}_l$, {$C^{\mu y}_l$,} $C^{\mu T}_l$ and
$C^{yT}_l$ allowed by the current observational constraints on the
magnetic fields, setting the amplitude of the
magnetic fields to the upper limit of
the constraints and choosing the cut-off scales appropriately.
The peak scales of the angular power spectra correspond to the cutoff scales of
the magnetic fields and the peak amplitudes of the auto-correlation spectra $C^{\mu\mu}_l$ and $C^{yy}_l$ are basically
determined by the total energy density of the decay of the magnetic fields. However, since the angular power spectra have the dependence on $l^2$,
the amplitude are suppressed on the observation scales.
On the other hand, we found that the cross correlation between 
the distortions $C^{\mu y}_l$ is small since different scales dominantly 
contributing to $\mu$- and $y$-distortions are different.
 The cross-correlation with CMB temperature and distortion anisotropies,
$C^{\mu T}_l$ and $C^{yT}_l$, are also suppressed more than
$C^{\mu\mu}_l$ and $C^{yy}_l$, since temperature fluctuations on such a
small length scale are exponentially suppressed due to Silk damping.

Following the numerical calculation of the angular power spectra, we have also discussed the possibility of detection of
anisotropic CMB distortions induced by primordial magnetic fields.
Although PIXIE is absolutely-calibrated experiment and have a high
sensitivity for the measurement of the CMB distortions, it is not able to measure
$C^{\mu\mu}_l$ and $C^{yy}_l$ since it can reach the small scale only up
to $l\sim 100$ in the current design.
Following the method proposed in \cite{Ganc:2012ae},
relatively calibrated experiments 
can also measure anisotropies of CMB distortions. LiteBIRD can detect $C^{yy}_l$ due to the magnetic fields
with the large tilt $n\gtrsim 3$ and the amplitude close
to the upper limit from current observations.
CMBpol can measure $C^{yy}_l$ by weaker magnetic fields.
 According to the situation, it might detect not $C^{yy}_l$ but $C^{\mu\mu}_l$.
Through observations of anisotropies of CMB distortions by these future
CMB satellites,
we might confirm existence of the primordial magnetic fields
with highly blue-tilted power spectrum or put the novel constraint on
such magnetic fields.
On the other hand, $C^{\mu T}$ and $C^{yT}$ are far from the detectable level in
both observation.
In small-scale CMB measurements by ground-based telescopes such as SPT,
it is difficult to search anisotropies of CMB distortions due to magnetic fields.
This is because the recent result of SPT is consistent with the SZ effect, 
which induces $y$-distortions indistinguishable from those by magnetic fields, 
and contributions from magnetic fields should be subdominant.
This leads to the upper limit on magnetic fields:
$B_0<1.8\times 10^2~{\rm nG}$ for $k_c=10~{\rm Mpc}^{-1}$, which is larger by
an $\mathcal{O}(1)$ factor than the COBE constraint,
Eq.~(\ref{COBEcons}).  The $\mu$ anisotropies are too small to be detected by
SPT.

We mention to another contribution from primordial magnetic fields to the cross-correlation between CMB distortion and temperature anisotropies.
In addition to the scalar passive mode considered here as the CMB temperature anisotropies sourced from the primordial magnetic fields,
primordial magnetic fields also induce the so-called compensated scalar magnetic mode of CMB temperature fluctuations \cite{Shaw:2009nf}.
Although it is subdominant compared with the scalar passive mode for $l\lesssim 5000$ \cite{Shaw:2009nf}, it becomes the dominant component for higher $l$.
This is because the scalar magnetic mode is actively produced and not suppressed exponentially like the scalar passive mode.
Since small-scale perturbations also contribute to $C^{X T}_l$ for small $l$,
it could be possible that the scalar magnetic mode makes a considerable contribution to $C^{X T}_l$ in the observable range
with the blue-tilted power spectrum of the magnetic fields.
As a brute-force estimation for such contribution,
let us consider the unrealistic case where the exponential suppression due to the Silk damping is absent in the temperature anisotropies.
In this case,
we can expect from FIG. 2 in \cite{Shaw:2009nf} that 
the temperature anisotropy induced from the scalar passive mode is larger than that generated by the scalar magnetic mode
by several orders of magnitude even on small scales.
Furthermore, the angular cross power spectra between the temperature anisotropy induced from the scalar passive mode
and the CMB distortions due to the primordial magnetic fields, $C^{X T}_l$, is expected to be comparable to the auto angular power spectrum
of the temperature anisotropy, $C^{T T}_l$, in the absence of Silk damping.
This is because Eqs. (\ref{CmumuB}) and (\ref{CmuTB}) (Eqs. (\ref{CyyB}) and (\ref{CyTB})) have almost same forms except for the difference in $\mathcal{O}(1)$ prefactors when we make an approximation that $\Delta^S_l(k)\sim j_l(kr_{\rm rec})$.
We therefore see that
$C^{XT}_l$ due to the scalar magnetic mode would be smaller than $C^{X X}_l$ by several orders of magnitudes.
Hence, when we consider the detectability of the CMB distortions due to the primordial magnetic fields in the previous section,
including $C^{XT}_l$ due to the scalar magnetic mode in the analysis seems not to increase the detectability of the distortions.
Then, we do not include the effects of the scalar magnetic mode in this paper
and would like to consider the detailed analysis including such effect as a future issue.

Although PIXIE will not detect anisotropic parts of $\mu$ and $y$
induced by primordial magnetic fields, it can detect their isotropic
parts if $\mu \gtrsim 5\times10^{-8}$ or $y \gtrsim 10^{-8}$, which
correspond to primordial magnetic fields with $\rho_B/\rho_\gamma\gtrsim
10^{-8}$ or $B_0\gtrsim 1{\rm nG}$.  Combining the result of PIXIE with
that of observations of $\mu$ and $y$ anisotropies by other satellites,
we might be able to confirm that the source of CMB distortions is
primordial magnetic fields.  Such a type of analysis will shed light on
physics in the early Universe in a novel way.

%----------------------------------%
\acknowledgments
%
%----------------------------------%

S.Y. would like to thank Jens Chluba for useful discussion.
K.M.,  S.Y. and T.S. would like to thank the Japan Society for the Promotion of Science for financial support.
H.T. is supported by the DOE at the Arizona State University.

\end{document}